\documentclass[12pt,draftclsnofoot,onecolumn]{IEEEtran}

\usepackage[latin1]{inputenc}
\usepackage{graphicx}
\usepackage{color}
\usepackage{placeins}
\usepackage{float}
\usepackage{tabularx,colortbl}
\usepackage{amssymb}
\usepackage{amsthm}
\usepackage{cite}
\usepackage{amsmath}
\usepackage{caption2}

\usepackage{cases,subeqnarray}
\usepackage{bm,multirow,bigstrut}
\usepackage{textcomp}
\usepackage{latexsym,bm}
\usepackage{booktabs,changebar}
\usepackage{xcolor}
\usepackage{mathtools}
\usepackage{dsfont}
\usepackage{extarrows}

\usepackage{subfigure}

\usepackage{algorithm}
\usepackage{algpseudocode}
\theoremstyle{plain}
\newtheorem{thm}{Theorem}

\theoremstyle{plain}
\newtheorem{rem}{Remark}
\newtheorem{cor}{Corollary}

\IEEEoverridecommandlockouts

\begin{document}

\title{UAV Communications with WPT-aided Cell-Free Massive MIMO Systems}

\author{Jiakang Zheng, Jiayi~Zhang,~\IEEEmembership{Senior Member,~IEEE}, \\and Bo Ai,~\IEEEmembership{Senior Member,~IEEE}

\thanks{A conference version of this paper has appeared in IEEE International Conference on Communications 2021 \cite{zheng2021analysis}.}

\thanks{J. Zheng and J. Zhang are with the School of Electronics and Information Engineering, Beijing Jiaotong University, Beijing 100044, P. R. China. (e-mail: \{20111047, jiayizhang\}@bjtu.edu.cn).}
\thanks{B. Ai is with the State Key Laboratory of Rail Traffic Control and Safety, Beijing Jiaotong University, Beijing 100044, China. (e-mail: boai@bjtu.edu.cn).}
}

\maketitle
\vspace{-1cm}
\begin{abstract}
Cell-free (CF) massive multiple-input multiple-output (MIMO) is a promising solution to provide uniform good performance for unmanned aerial vehicle (UAV) communications. In this paper, we propose the UAV communication with wireless power transfer (WPT) aided CF massive MIMO systems, where the harvested energy (HE) from the downlink WPT is used to support both uplink data and pilot transmission. We derive novel closed-form downlink HE and uplink spectral efficiency (SE) expressions that take hardware impairments of UAV into account. UAV communications with current small cell (SC) and cellular massive MIMO enabled WPT systems are also considered for comparison. It is significant to show that CF massive MIMO achieves two and five times higher 95\%-likely uplink SE than the ones of SC and cellular massive MIMO, respectively. Besides, the large-scale fading decoding receiver cooperation can reduce the interference of the terrestrial user. Moreover, the maximum SE can be achieved by changing the time-splitting fraction. We prove that the optimal time-splitting fraction for maximum SE is determined by the number of antennas, altitude and hardware quality factor of UAVs. Furthermore, we propose three UAV trajectory design schemes to improve the SE. It is interesting that the angle search scheme performs best than both AP search and line path schemes. Finally, simulation results are presented to validate the accuracy of our expressions.
\end{abstract}

\begin{IEEEkeywords}
Cell-free massive MIMO, unmanned aerial vehicle, wireless power transfer, hardware impairments, harvested energy, spectral efficiency.
\end{IEEEkeywords}

\IEEEpeerreviewmaketitle

\section{Introduction}
The demand for data throughput has been rapidly growing for decades due to a large amount of user equipments (UEs) and various wireless applications \cite{wong2017key,ai20205g}. In order to fullfill this demand, cell-free (CF) massive multiple-input multiple-output (MIMO) has been recently proposed as a promising technology for providing a higher and more uniform spectral efficiency (SE) to the UEs in wireless networks \cite{zhang2020prospective,Ngo2017Cell}. CF massive MIMO systems consist of many geographically distributed access points (APs) connected to a central processing unit (CPU) for coherently serving the UEs by spatial multiplexing on the same time-frequency resource \cite{Ngo2017Cell,bjornson2019making}.

Compared with small-cell (SC) systems and cellular massive MIMO systems, there are no cell boundaries and many more APs than UEs in CF massive MIMO systems. The APs only serve UEs within their own cell in conventional SC and cellular massive MIMO systems, which may cause large inter-cell interference. The results in \cite{Ngo2017Cell,bjornson2019making} showed that the CF massive MIMO system provides better performance than SC and cellular massive MIMO systems in terms of 95\%-likely per-user SE. Based on these seminal works, a plethora of papers of CF massive MIMO have been published in recent years. For instance, the performance of CF massive MIMO system with Rician fading channels was investigated in \cite{ozdogan2019performance}. A thorough investigation of the channel hardening and favorable propagation phenomena in CF massive MIMO systems from a stochastic geometry perspective was provided in \cite{chen2018channel}. A practical framework based on deep learning was proposed in \cite{jin2019channel} to perform channel estimation in CF millimeter-wave massive MIMO systems. The authors in \cite{zheng2020efficient} investigated the effect of hardware impairments on the performance of CF massive MIMO systems and point out that hardware impairments at the transmitter has larger effect on the uplink average SE compared with the one at the receiver.

Recently, there are several important and practical applications of unmanned aerial vehicles (UAV), such as mobile base stations, mobile relays, and mobile data collections \cite{9043712}. In addition, UAVs have been found many promising applications, such as aerial inspection, photography, precision agriculture, traffic control, search and rescue, package delivery, and telecommunications \cite{zeng2019accessing}.
In addition, there are many significant advantages for UAV-aided wireless communications \cite{zeng2016wireless,zeng2018cellular}. For examples, UAVs can be deployed rapidly as aerial base stations or aerial mobile relays to provide improved performance for existing wireless communication networks and to support emergent service in disaster areas \cite{shen2020multi}. Furthermore, UAVs are useful for data collection and dissemination in wireless sensor networks \cite{zhan2019completion}. Many important and fundamental aspects of UAV-aided wireless communications have been studied in last decade. The authors in \cite{mozaffari2018beyond} introduced a novel concept of three-dimensional (3D) cellular networks and proposed latency-optimal cell association to improve the SE. The authors in \cite{mei2019cellular} proposed new inter-cell interference coordination designs to mitigate the strong uplink interference in cellular-connected UAV communication. By leveraging the use of UAVs for data offloading, the authors in \cite{lyu2018uav} provided a new hybrid network architecture for cellular systems. However, UAV communication with CF massive MIMO systems is rarely investigated. The numerical results in \cite{d2020analysis,d2019cell} revealed that, in UAV-aided wireless communications, CF massive MIMO may provide better performance than a traditional cellular massive MIMO network.
In addition, an UAV is used as the mobile base station (BS) to further improve the performance of the CF massive MIMO system \cite{9336017}.

Although with promising benefits, wireless communications with UAVs are also faced with several challenges \cite{zeng2016wireless,gupta2015survey,mozaffari2019tutorial}. The performance and operational duration of a UAV system is fundamentally constrained by the limited onboard energy \cite{zeng2016wireless}. More specifically, energy consumption of the UAVs can be broadly classified into mobility energy consumption and communication energy consumption, which aim to satisfy the movement of the UAVs and the communication requirement, respectively \cite{liu2018energy}. Using radio frequency energy harvesting to achieve wireless power transfer (WPT) and wireless information transfer has been proposed as an effective method to replenish the energy of UAV communication requirement \cite{wang2019coverage}. The authors in \cite{xie2018throughput} investigated the throughput of an UAV-enabled wireless powered communication network with downlink WPT and uplink wireless information transfer. An UAV enabled mobile edge computing wireless powered system was studied in \cite{zhou2018computation} to maximize the achievable computation rate.
Moreover, the authors in \cite{9075988} studied the robust joint design for an energy-constrained UAV secure communication system with WPT to maximize the minimum secrecy rate.
Another main challenge stems from the size, weight, and power constraints of UAVs, which could limit their communication, computation, and endurance capabilities \cite{zeng2016wireless}. Therefore, the UAV hardware impairment acts as a non-linear filter in practice, because of limited resolutions of analog-to-digital and digital-to-analog converters, power amplifier and oscillator phase noise \cite{hou2020hardware}. However, most of works consider the residual hardware impairments model, which is uncorrelated with input signal \cite{hou2020hardware,li2020uav}. Considering the hardware impairments correlated with input signal, the authors in \cite{bjornson2017massive} proposed a more realistic hardware impairments model.

Different from conventional wireless systems in a state of stillness, UAV-enabled wireless communications require proper trajectory design schemes due to energy constraints \cite{mozaffari2017mobile}. Therefore, trajectory design is one important aspect for the success of UAV-enabled wireless communications \cite{zeng2018trajectory}. The authors in \cite{sun2019optimal} studied the joint design of the 3D aerial trajectory and the wireless resource allocation for maximization of the system sum throughput. The authors in \cite{zhang2019securing} and \cite{cui2018robust} exploited the mobility of UAV via its trajectory design to tackle the information security in the physical layer. A flight time minimization problem was solved in \cite{gong2018flight}, considering a scenario where an UAV collects data from a set of sensors on a straight line. The authors in \cite{yang2018energy} investigated the optimal energy trade-off between the UAV and its served ground terminal via UAV trajectory designs. A general UAV-enabled radio access network was studied in \cite{zhang2018uav}, while each periodic flight duration of the UAV and the mission completion time for saving UAV time were minimized via optimizing the UAV trajectory in periodic and one-time operation scenarios, respectively. Moreover, the authors in \cite{xu2018uav} exploited the mobility of the UAV to maximize the energy transferred to all energy receivers by optimizing the UAV trajectory in an UAV-enabled multiuser WPT system.

Motivated by the aforementioned observations, we study the UAV communication with CF massive MIMO enabled WPT systems, where the harvested energy (HE) from the downlink WPT is used to support the uplink data and pilot transmission. In addition, the hardware impairments effect at UAV is also considered. We investigate the uplink energy harvesting and downlink SE performance of the considered systems. The performance of the corresponding SC system and cellular massive MIMO are analyzed for comparison. Moreover, a useful angle search trajectory design scheme is proposed to improve the SE. The heuristic AP search and line path trajectory design schemes are also considered for comparison. The specific contributions of our work are listed as follows:
\begin{itemize}
  \item We first derive closed-form expressions for the downlink HE and uplink SE of the UAV communication with CF massive MIMO enabled WPT systems taking into account a realistic hardware impairment model at UAV. Our results show that CF massive MIMO performs better than cellular massive MIMO both in terms of downlink HE and uplink SE. For a fair comparison, we multiply the downlink transmit power by the number of APs, but the uplink SE of SC is still smaller than CF massive MIMO.
  \item We find that the maximum SE can be observed by changing the ratio of downlink WPT and uplink data transmission. Increasing the number of antennas and decreasing the altitude of UAV both can improve the uplink SE and lead to longer uplink data transmission are preferred for the optimal operating point of SE. It is also found that hardware impairment of UAV has a bad effect on the SE and reduces the optimal downlink WPT time which we obtain maximum SE. Moreover, the large-scale fading decoding (LSFD) receiver cooperation can reduce the interference of the terrestrial UE (TUE).
  \item We investigate the SE-improved UAV trajectory design taking into account a destination. Angle search trajectory design scheme is proposed to improve the SE. Compared with AP search and line path trajectory design schemes, we find that angle search trajectory design scheme performs better in CF massive MIMO. In addition, the angle search scheme can bypass the APs with large interference in both CF massive MIMO and SC systems, but can not in cellular massive MIMO systems.
\end{itemize}
%

The rest of this paper is organized as follows.
In Section \ref{se:model}, we model the channel mode, uplink channel estimation, downlink energy harvesting and uplink data transmission of CF massive MIMO.
In Section \ref{se:performance}, we analyze the downlink HE and uplink SE of the UAV-aided wireless communication enabled WPT systems with CF massive MIMO, SC and cellular massive MIMO architectures, respectively.
In Section \ref{se:trajectory}, we provide angle search and AP search trajectory design schemes to achieve performance gain, and line path trajectory design scheme is considered for comparison.
The specific simulation results are given in Section \ref{se:numerical}.
Finally, we summarize the full text in Section \ref{se:conclusion}.

\textit{Notations:} Column vectors and matrices are represented by boldface lowercase letters $\mathbf{x}$ and boldface uppercase letters $\mathbf{X}$, respectively. We use superscripts $x^\mathrm{*}$ and $\mathbf{x}^\mathrm{H}$ to \mbox{represent} conjugate and conjugate transpose, respectively. ${{\mathbf{I}}_N}$ is the $N \times N$ identity matrix, and ${\log _2}\left(  \cdot  \right)$ denotes the logarithm with base 2.
The Euclidean norm, the expectation operators and the definitions are denoted by $\left\|  \cdot  \right\|$, $\mathbb{E}\left\{  \cdot  \right\}$, and $\triangleq$, respectively.
Finally, $x \sim \mathcal{C}\mathcal{N}\left( {0,{\sigma^2}} \right)$ represents a circularly symmetric complex Gaussian random variable $x$ with variance $\sigma^2$.

\vspace{-2mm}

\section{System Model}\label{se:model}

\begin{figure}[t]
\centering
\includegraphics[scale=0.6]{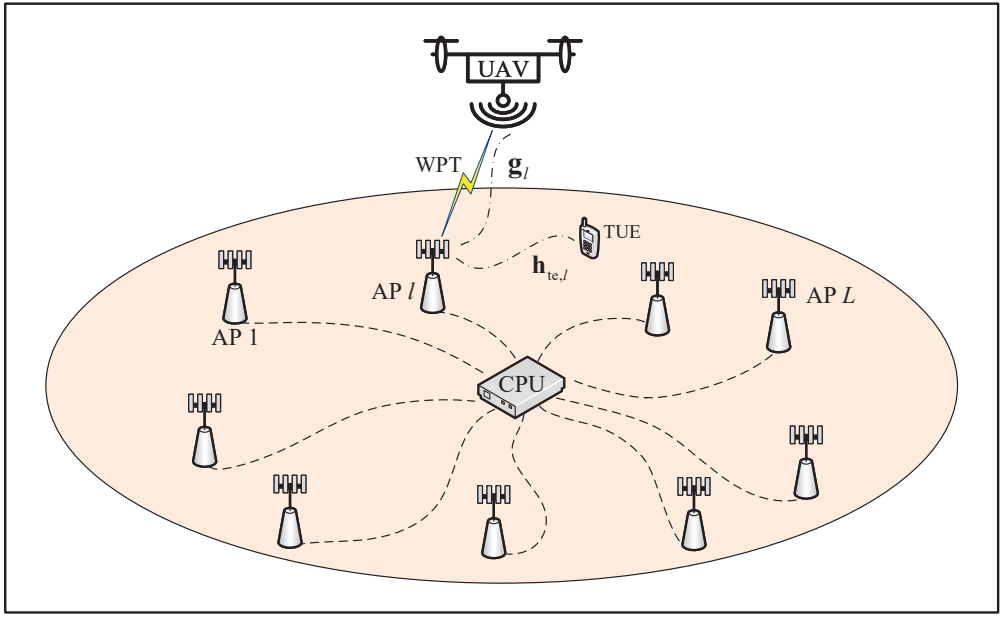}
\caption{UAV communications with CF massive MIMO enabled WPT systems.} \vspace{-4mm}
\label{system_model}
\end{figure}

\begin{figure}[t]
\centering
\includegraphics[scale=0.95]{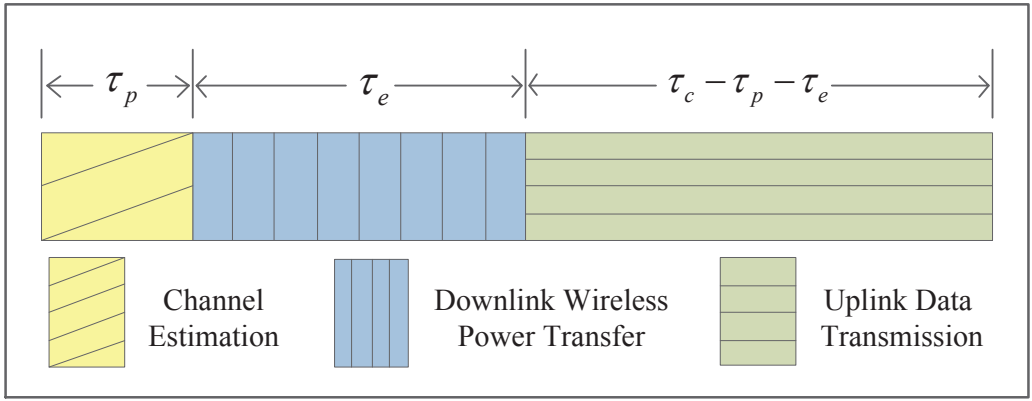}
\caption{Coherence block structure} \vspace{-4mm}
\label{block}
\end{figure}

As illustrated in Fig.~\ref{system_model}, we consider a wireless powered CF massive MIMO system consisting of $L$ APs and one energy-harvesting UAV. Each AP is equipped with $N$ antennas and UAV is equipped with a single antenna\footnote{Large signal processing complexity as well as the power consumption make it quite costly to employ multiple antennas in UAVs.}. The APs are connected to a CPU via fronthaul links. We assume that all $L$ APs simultaneously serve the UAV on the same time-frequency resource, and the hardware impairment at UAV acts as a non-linear filter \cite{bjornson2017massive}. Besides, the effect of TUE interference will be investigated in Section \ref{interference}.

We consider frame-based transmissions over flat-fading channels on a single frequency band. The communication is divided into coherence blocks consisting of $\tau_c$ channel uses as shown in Fig.~\ref{block}. We assume that the uplink training phase occupies $\tau_p$ channel uses for channel estimation, the downlink WPT phase occupies $\tau_e$ channel uses and uplink data transmission phase occupies $\tau_c-\tau_p-\tau_e$ channel uses.
We define $\rho  = {\tau _e}/\left( {{\tau _c} - {\tau _p}} \right)$ as time-splitting fraction, which reflects the proportion of downlink energy harvesting and uplink data transmission.

\subsection{Propagation Model}

A 3D Euclidean space is used to determine the locations of $L$ APs and the UAV. We assume horizontal plane coordinates of APs are ${{\mathbf{q}}_l} \triangleq \left[ {{x_l},{y_l}} \right],l = 1, \ldots ,L$ and the horizontal plane coordinate of UAV is ${{\mathbf{q}}_{\text{u}}} \triangleq \left[ {{x_{\text{u}}},{y_{\text{u}}}} \right]$. In addition, we assume the high of APs are zero and the UAV flies at a fixed altitude $H$.
Considering the UAV trajectory, we define the coherence block as time slot, which is the minimum flight time interval. At each time slot, the UAV does channel estimation, downlink wireless energy harvesting and uplink data transmission. Then, the position of UAV at the $n$th time slot can be expressed as
\begin{align}
{{\mathbf{q}}_{\text{u}}}\left[ n \right] \triangleq \left[ {{x_{\text{u}}}\left[ n \right],{y_{\text{u}}}\left[ n \right]} \right],n = 0,1, \ldots ,{N_{{\text{slot}}}}.
\end{align}
Due to the high probability of line-of-sight (LoS) link in UAV communication, the large-scale fading between the UAV and the AP $l$ can be modeled by the free-space path loss as
\begin{align}
{\zeta _l}\left[ n \right] = \frac{{{\beta _0}}}{{{{\left\| {{{\mathbf{q}}_{\text{u}}}\left[ n \right] - {{\mathbf{q}}_l}} \right\|}^2} + {H^2}}}, l = 1, \ldots ,L,n = 0,1, \ldots ,{N_{{\text{slot}}}},
\end{align}
where ${\beta _0}$ denotes the received power at the reference distance (e.g., $d = 1$ m) between the transmitter and the receiver.

Take the impact of buildings and obstructions into account, we need to use more refined channel models to reflect the change of propagation environment.
For the small-scale fading of UAV communications, we use spatially correlated and altitude dependent Rician fading channels, which include a probabilistic LoS component on top of a Rayleigh distributed component modeling the scattered multipath.
Therefore, multiplying the large-scale fading by the small-scale fading effect, the channel gain between UAV and AP $l$ at time slot $n$ is expressed as
\begin{align}
{{\mathbf{g}}_l}\left[ n \right] \sim \mathcal{C}\mathcal{N}\left( {{{{\mathbf{\bar h}}}_l}\left[ n \right],{{\mathbf{R}}_l}}\left[ n \right] \right),l = 1, \ldots ,L,n = 0,1, \ldots ,N_\text{slot},
\end{align}
where ${{{\mathbf{\bar h}}}_l}\left[ n \right] \in {\mathbb{C}^N}$ is the LoS component, and ${{\mathbf{R}}_l}\left[ n \right] \in {\mathbb{C}^{N \times N}}$ is the positive semi-definite covariance matrix describing the spatial correlation of the non-line-of-sight (NLoS) components.
It is worth noting that channel gain also can be expressed as ${{\mathbf{g}}_l}\left[ n \right] = {{{\mathbf{\bar h}}}_l}\left[ n \right] + {{\mathbf{h}}_l}\left[ n \right]$, where ${{\mathbf{h}}_l}\left[ n \right] \sim \mathcal{C}\mathcal{N}\left( {{\mathbf{0}},{{\mathbf{R}}_l}\left[ n \right]} \right)$ is a Rayleigh distributed component.

\subsection{Uplink Channel Estimation}

Due to one UAV is considered, we assume that the UAV only send one pilot ($\tau_p=1$) to each AP for uplink channel estimation. Then, the received signal in AP $l$ at time slot $n$ is
\begin{align}
{{\mathbf{z}}_l}\left[ n \right] &= {{\mathbf{g}}_l}\left[ n \right]\left( {\sqrt {\kappa p\left[ n \right]}  + \eta\left[ n \right] } \right) + {{\mathbf{n}}_l}\left[ n \right] \notag\\
&=\sqrt {\kappa p\left[ n \right]} {{{\mathbf{\bar h}}}_l}\left[ n \right] + \sqrt {\kappa p\left[ n \right]} {{\mathbf{h}}_l}\left[ n \right] + \eta\left[ n \right] {{{\mathbf{\bar h}}}_l}\left[ n \right] + \eta\left[ n \right] {{\mathbf{h}}_l}\left[ n \right] + {{\mathbf{n}}_l}\left[ n \right],
\end{align}
where $p\left[ n \right] \geqslant 0$ is the pilot transmit power of UAV and ${{\mathbf{n}}_l}\left[ n \right] \sim \mathcal{C}\mathcal{N}\left( {{\mathbf{0}},{\sigma ^2}{{\mathbf{I}}_N}} \right)$ is the receiver noise in AP $l$ at time slot $n$. In addition, $0 \leqslant \kappa \leqslant 1$ is the hardware quality factor of the UAV, and $\eta\left[ n \right] \sim \mathcal{C}\mathcal{N}\left( {0,\left( {1 - \kappa } \right)p\left[ n \right]} \right)$ is the hardware impairment of the UAV.
Using linear minimum mean square error (LMMSE) estimation, each AP can compute the LMMSE estimate ${{{\mathbf{\hat g}}}_l}\left[ n \right]$ of the channel coefficient ${{{\mathbf{ g}}}_l}\left[ n \right]$ as
\begin{align}
{{{\mathbf{\hat g}}}_l}\left[ n \right] = {{{\mathbf{\bar h}}}_l}\left[ n \right] + \sqrt {\kappa p\left[ n \right]} {{\mathbf{R}}_l}\left[ n \right]{{\mathbf{\Psi }}_l}\left[ n \right]\left( {{{\mathbf{z}}_l}\left[ n \right] - {{{\mathbf{\bar z}}}_l}\left[ n \right]} \right),
\end{align}
where
\begin{align}
{{{\mathbf{\bar z}}}_l}\left[ n \right] &= \sqrt {\kappa p\left[ n \right]} {{{\mathbf{\bar h}}}_l}\left[ n \right], \\
{{\mathbf{\Psi }}_l}\left[ n \right] &= {\left( {p\left[ n \right]{{\mathbf{R}}_l}\left[ n \right] + \left( {1 - \kappa } \right)p\left[ n \right]{{{\mathbf{\bar h}}}_l}\left[ n \right]{\mathbf{\bar h}}_l^{\text{H}}\left[ n \right] + {\sigma ^2}{{\mathbf{I}}_N}} \right)^{ - 1}}.
\end{align}
The estimate ${{{\mathbf{\hat g}}}_l}\left[ n \right]$ and the estimation error ${{{\mathbf{\tilde g}}}_l}\left[ n \right] = {{\mathbf{g}}_l}\left[ n \right] - {{{\mathbf{\hat g}}}_l}\left[ n \right]$ are distributed as $\mathcal{C}\mathcal{N}\left( {{{\mathbf{\bar h}}}\left[ n \right],{{\mathbf{Q}}_l}\left[ n \right]} \right)$ and $\mathcal{C}\mathcal{N}\left( {{\mathbf{0}},{{\mathbf{C}}_l}\left[ n \right]} \right)$, where
\begin{align}\label{Q}
{{\mathbf{Q}}_l}\left[ n \right] = {{\mathbf{R}}_l}\left[ n \right]-{{\mathbf{C}}_l}\left[ n \right] = \kappa p\left[ n \right] {{\mathbf{R}}_l}\left[ n \right]{{\mathbf{\Psi }}_l}\left[ n \right]{{\mathbf{R}}_l}\left[ n \right].
\end{align}

\subsection{Downlink Energy Harvesting}

Using the beamforming scalar ${\mathbf{w}}\left[ n \right] = {{{{{\mathbf{\hat g}}}_l}}\left[ n \right]}/{{\sqrt {\mathbb{E}\left\{ {{{\left\| {{{{\mathbf{\hat g}}}_l}\left[ n \right]} \right\|}^2}} \right\}} }}$, the transmitted signal from AP $l$ at time slot $n$ is
\begin{align}
{{\mathbf{x}}_l^\text{cf}}\left[ n \right] = \sqrt {{p_{\text{d}}^\text{cf}}} {\mathbf{w}}\left[ n \right] v_l \left[ n \right] ,
\end{align}
where $p_\text{d}^\text{cf}$ is the downlink transmission power, and $v_l\left[ n \right] \sim \mathcal{C}\mathcal{N}\left( {0,1} \right)$ is different energy symbols sent to the UAV at time slot $n$.
Then, the received signal at the UAV is expressed as
\begin{align}\label{r_received}
r^\text{cf}\left[ n \right] = \sqrt {\kappa {p_{\text{d}}^\text{cf}}} \sum\limits_{l = 1}^L {{\mathbf{g}}_l^{\text{H}}\left[ n \right]{\mathbf{w}}\left[ n \right]} v_l\left[ n \right]  + \mu\left[ n \right]  + n\left[ n \right] ,
\end{align}
where $\mu\left[ n \right]$ is the hardware impairment of UAV, which is distributed as $\mu\left[ n \right]  \sim \mathcal{C}\mathcal{N}\left( {0,\left( {1 - \kappa } \right){p_{{\text{in}}}^\text{cf}}\left[ n \right]} \right)$. $p_\text{in}^\text{cf}\left[ n \right]$ is the received power in UAV at time slot $n$ expressed as\footnote{It is worth noting that the received power is only used for the communication power of UAV, which includes the information signal power and the constant circuit power caused by hardware impairment. In addition, the flight power of UAVs could come from the battery or solar energy \cite{sun2019optimal}.}
\begin{align}\label{P_in}
{p_{{\text{in}}}^\text{cf}}\left[ n \right] =  \sum\limits_{l = 1}^L {\mathbb{E}\left\{ {{{\left| {\sqrt {{p_{\text{d}}^\text{cf}}} {\mathbf{g}}_l^{\text{H}}\left[ n \right]{\mathbf{w}}\left[ n \right]} \right|}^2}} \right\}} .
\end{align}
We assume the energy due to the hardware impairment and the noise in \eqref{r_received} cannot be harvested. Therefore, the HE by UAV is ${P_{{\text{HE}}}^\text{cf}}\left[ n \right] = \frac{\tau_e}{\tau_c} \kappa {p_{{\text{in}}}^\text{cf}}\left[ n \right]$. Note that the energy used for pilot transmission is drawn from ${P_{{\text{HE}}}^\text{cf}}\left[ n -1 \right]$, which is the HE in the last time slot\footnote{The UAV has an initial energy $p\left[ 0 \right]=1$ dBm to send pilot for connecting to the wireless powered networks.}.
At steady state, we assume a fraction $\partial = \tau_p/\left(\tau_c-\tau_e\right)$ of the HE ${P_{{\text{HE}}}^\text{cf}}\left[ n \right]$ is used by UAV to send pilot at the next time slot, and the left HE $\left(1-\partial\right){P_{{\text{HE}}}^\text{cf}}\left[ n \right]$ is used for uplink data transmission.

\subsection{Uplink Data Transmission}

During the uplink data transmission, the received complex baseband signal in AP $l$ at time slot $n$ is
\begin{align}
{{\mathbf{y}}_l^\text{cf}}\left[ n \right] = {{\mathbf{g}}_l}\left[ n \right]\left( {\sqrt {\kappa {p_{\text{u}}^\text{cf}\left[ n \right]}} s\left[ n \right] + \eta\left[ n \right] } \right) + {{\mathbf{n}}_l}\left[ n \right]
\end{align}
where ${{p_{\text{u}}^\text{cf}}}\left[ n \right]$ is the uplink data transmission power, and $\eta\left[ n \right]  \sim \mathcal{C}\mathcal{N}\left( {0,\left( {1 - \kappa } \right){p_{\text{u}}^\text{cf}}\left[ n \right]} \right)$ is the hardware impairment of the UAV. ${{\mathbf{n}}_l}\left[ n \right] \sim \mathcal{C}\mathcal{N}\left( {0,{\sigma ^2}{{\mathbf{I}}_N}} \right)$ is the receiver noise.

To detect the symbol transmitted from the UAV, the $l$th AP multiplies the received signal ${{\mathbf{y}}_l^\text{cf}}\left[ n \right]$ with the conjugate of its (locally obtained) channel estimate as
\begin{align}\label{s_hat_l}
{{\overset{\lower0.5em\hbox{$\smash{\scriptscriptstyle\smile}$}}{s} }_l}\left[ n \right] = {\mathbf{\hat g}}_l^{\text{H}}\left[ n \right]{{\mathbf{y}}_l^\text{cf}}\left[ n \right] = {\mathbf{\hat g}}_l^{\text{H}}\left[ n \right]\left( {{{\mathbf{g}}_l}\left[ n \right]\left( {\sqrt {\kappa {p_{\text{u}}^\text{cf}\left[ n \right]}} s\left[ n \right] + \eta\left[ n \right] } \right) + {{\mathbf{n}}_l}\left[ n \right]} \right).
\end{align}

\begin{rem}
During the UAV movement, the channel estimation, the beamforming scalar and MR combing vector change in real time. TABLE \ref{table_compute} summarizes the computational complexity in terms of complex multiplications per coherence block for different communication phases.
\end{rem}

\begin{table}[t]
\centering
\newcommand{\tabincell}[2]{\begin{tabular}{@{}#1@{}}#2\end{tabular}}
\caption{Computational complexity per coherence block.}
\setlength{\tabcolsep}{3.4mm}{
\vspace{4mm}
\centering
\begin{tabular}{|c|c|c|}
    \hline
    \hline
     Communication Phase & Operation & Complex Multiplications \cr\hline
    \multirow{2}{*}{Channel Estimation}&  Computing the precomputed statistical matrix  & $KL \left(4N^3-N\right)/3$ \cr\cline{2-3}
    & \tabincell{c}{Multiplying the precomputed statistical matrix  \\ with the received signal vector } & $KL N^2$ \cr\hline
    \multirow{2}{*}{Downlink Transmission}&  Computing the beamforming scalar vectors &  $KLN$ \cr\cline{2-3}
    & \tabincell{c}{Multiplying beamforming scalar vectors with \\ the transmit energy symbols} & $KL \tau_e N$ \cr\hline
    \multirow{2}{*}{Uplink Reception}& Computing MR combing vectors  & ------ \cr\cline{2-3}
    & \tabincell{c}{Multiplying MR combing vectors \\ with the received signal} & $KL \left(\tau_c-\tau_p-\tau_e \right) N$ \cr\hline
    \hline
\end{tabular}}
\label{table_compute}
\end{table}

\section{Performance Analysis}\label{se:performance}

In this section, we investigate the downlink HE of UAV and uplink performance of CF massive MIMO systems within each time slot. Meanwhile, we consider SC and cellular massive MIMO systems for comparison. Novel expressions for the downlink HE and uplink SE are derived for both systems.

\subsection{CF massive MIMO}

\subsubsection{Downlink HE}

Using the beamforming scalar ${\mathbf{w}}\left[ n \right] = {{{{{\mathbf{\hat g}}}_l}}\left[ n \right]}/{{\sqrt {\mathbb{E}\left\{ {{{\left\| {{{{\mathbf{\hat g}}}_l}\left[ n \right]} \right\|}^2}} \right\}} }}$, and assume the energy due to hardware impairment and the noise in \eqref{r_received} cannot be harvested. The downlink HE is derived in the following theorem.
\begin{thm}\label{thm1}
Based on \eqref{P_in}, we can derive the downlink HE by UAV as
\begin{align}
{P_{{\mathrm{HE}}}^\mathrm{cf}}\left[ n \right] = \frac{\tau_e}{\tau_c} \kappa {p_{{\mathrm{in}}}^\mathrm{cf}}\left[ n \right] = \frac{\tau_e}{\tau_c} \kappa \left( { {p_{\mathrm{d}}^\mathrm{cf}}\sum\limits_{l = 1}^L {\frac{{ {\Upsilon _l}\left[ n \right]  + {\left( {\mathrm{tr}}\left( {{{\mathbf{Q}}_l}}\left[ n \right] \right) + {{\left\| {{{{\mathbf{\bar h}}}_l}}\left[ n \right] \right\|}^2} \right)^2} }}{{{\mathrm{tr}}\left( {{{\mathbf{Q}}_l}}\left[ n \right] \right) + {{\left\| {{{{\mathbf{\bar h}}}_l}}\left[ n \right] \right\|}^2}}}}  } \right),
\end{align}
where
\begin{align}
{\Upsilon _l}\left[ n \right] = {\mathrm{tr}}\left( {{{\mathbf{R}}_l}\left[ n \right]{{\mathbf{Q}}_l}\left[ n \right]} \right) + {\mathbf{\bar h}}_l^{\mathrm{H}}\left[ n \right]{{\mathbf{Q}}_l}\left[ n \right]{{{\mathbf{\bar h}}}_l}\left[ n \right] + {\mathbf{\bar h}}_l^{\mathrm{H}}\left[ n \right]{{\mathbf{R}}_l}\left[ n \right]{{{\mathbf{\bar h}}}_l}\left[ n \right].
\end{align}
\end{thm}
\begin{IEEEproof}
Please refer to Appendix A.
\end{IEEEproof}
Note that the downlink HE at the time slot $n$ is divided into two parts by the fraction $\partial$. Therefore, the uplink data transmission power is $p_\text{u}^\text{cf} \left[ n \right] = \frac{\tau_c}{\tau_c-\tau_p-\tau_e} \left(1-\partial\right){P_{{\text{HE}}}^\text{cf}}\left[ n \right]$, and the next slot pilot transmit power is $p \left[ n+1 \right] = \frac{\tau_c}{\tau_p} \partial {P_{{\text{HE}}}^\text{cf}}\left[ n \right]$. Then, we can find that $p_\text{u}^\text{cf} \left[ n \right]=p \left[ n+1 \right]$.

\subsubsection{Uplink SE}

In CF massive MIMO systems, the obtained quantity \eqref{s_hat_l} is sent to the CPU via the fronthaul as
\begin{align}\label{CPU}
\hat s \left[ n \right] &= \sum\limits_{l = 1}^L {{{\overset{\lower0.5em\hbox{$\smash{\scriptscriptstyle\smile}$}}{s} }_l}}\left[ n \right] = \underbrace {\sqrt {\kappa {p_{\text{u}}^\text{cf}}\left[ n \right]} \sum\limits_{l = 1}^L {\mathbb{E}\left\{ {{\mathbf{\hat g}}_l^{\text{H}}\left[ n \right]{{\mathbf{g}}_l}\left[ n \right]} \right\}s\left[ n \right]} }_{{\text{DS}\left[ n \right]}} + \underbrace {\sum\limits_{l = 1}^L {{\mathbf{\hat g}}_l^{\text{H}}\left[ n \right]{{\mathbf{g}}_l}\left[ n \right]\eta\left[ n \right] } }_{{\text{HI}\left[ n \right]}} \notag\\
&+ \underbrace {\sqrt {\kappa {p_{\text{u}}^\text{cf}}\left[ n \right]} \sum\limits_{l = 1}^L {\left( {{\mathbf{\hat g}}_l^{\text{H}}\left[ n \right]{{\mathbf{g}}_l}\left[ n \right] - \mathbb{E}\left\{ {{\mathbf{\hat g}}_l^{\text{H}}\left[ n \right]{{\mathbf{g}}_l}\left[ n \right]} \right\}} \right)s\left[ n \right]} }_{{\text{BU}\left[ n \right]}} + \underbrace {\sum\limits_{l = 1}^L {{\mathbf{\hat g}}_l^{\text{H}}\left[ n \right]{{\mathbf{n}}_l}\left[ n \right]} }_{{\text{NS}\left[ n \right]}},
\end{align}
where $\text{DS}\left[ n \right]$ represents the desired signal, $\text{BU}\left[ n \right]$ represents the beamforming gain uncertainty, $\text{HI}\left[ n \right]$ represents the hardware impairment effect, and $\text{NS}\left[ n \right]$ represents the noise term, respectively.
\begin{thm}\label{thm2}
A lower bound on the capacity of UAV is
\begin{align}\label{SE}
{\mathrm{SE}^\mathrm{cf}\left[ n \right]} = \frac{\tau_c-\tau_p-\tau_e}{\tau_c}\log \left( {1 + {\mathrm{SINR}^\mathrm{cf}\left[ n \right]}} \right),
\end{align}
where ${\mathrm{SINR}^{\mathrm{cf}}\left[ n \right]}$ is given by
\begin{align}
\frac{{\kappa {p_{\mathrm{u}}^\mathrm{cf}\left[ n \right]}{{\left| {\sum\limits_{l = 1}^L {\left( {{\mathrm{tr}}\left( {{{\mathbf{Q}}_l}}\left[ n \right] \right) + {{\left\| {{{{\mathbf{\bar h}}}_l}}\left[ n \right] \right\|}^2}} \right)} } \right|}^2}}}{{{p_{\mathrm{u}}^\mathrm{cf}\!\left[ n \right]}\!\sum\limits_{l = 1}^L {{\Upsilon _l}\!\left[ n \right]}  \!+\! \left( {1 \!-\! \kappa } \right){p_{\mathrm{u}}^\mathrm{cf}\!\left[ n \right]}{{\left( {\sum\limits_{l = 1}^L {\left( {{\mathrm{tr}}\left( {{{\mathbf{Q}}_l}}\!\left[ n \right] \right) \!+\! {{\left\| {{{{\mathbf{\bar h}}}_l}}\!\left[ n \right] \right\|}^2}} \right)} } \!\right)}^2} \!\!+\! {\sigma ^2}\sum\limits_{l = 1}^L {\!\left( {{\mathrm{tr}}\left( {{{\mathbf{Q}}_l}}\!\left[ n \right] \right) \!+\! {{\left\| {{{{\mathbf{\bar h}}}_l}}\!\left[ n \right] \right\|}^2}} \right)} }} .
\end{align}
\end{thm}
\begin{IEEEproof}
Please refer to Appendix B.
\end{IEEEproof}

\subsection{Small cell}

\subsubsection{Downlink HE}

In SC systems, the UAV harvests energy from the energy-maximizing AP. Then, the received signal from AP $l$ is
\begin{align}\label{r_l}
r_l^\text{sc}\left[ n \right] = \sqrt {\kappa {p_{\text{d}}^\text{cf}}} {\mathbf{g}}_l^{\text{H}}\left[ n \right]{{\mathbf{w}}_l}\left[ n \right]{v_l}\left[ n \right] + \mu \left[ n \right] + n\left[ n \right],
\end{align}
where $\mu\left[ n \right]  \sim \mathcal{C}\mathcal{N}\left( {0,\left( {1 - \kappa } \right){p_{{l,\text{in}}}^\text{sc}}\left[ n \right]} \right)$ is the hardware impairment of UAV. ${p_{{l,\text{in}}}^\text{sc}}\left[ n \right]$ is the received power from AP $l$ expressed as
\begin{align}
{p_{l,{\text{in}}}^\text{sc}}\left[ n \right] = \mathbb{E}\left\{ {{{\left| {\sqrt {{p_{\text{d}}^\text{sc}}} {\mathbf{g}}_l^{\text{H}}\left[ n \right]{{\mathbf{w}}_l}\left[ n \right]} \right|}^2}} \right\}.
\end{align}
\begin{cor}
Considering the energy due to hardware impairment and the noise in \eqref{r_l} cannot be harvested and using the beamformer ${\mathbf{w}}_l\left[ n \right] = {{{{{\mathbf{\hat g}}}_l}}\left[ n \right]}/{{\sqrt {\mathbb{E}\left\{ {{{\left\| {{{{\mathbf{\hat g}}}_l}\left[ n \right]} \right\|}^2}} \right\}} }}$, the HE by UAV at time slot $n$ can be derived as
\begin{align}
{{P}_{{\mathrm{HE}}}^\mathrm{sc}}\left[ n \right] \!=\! \frac{{{\tau _e}}}{{{\tau _c}}}\kappa \mathop {\max }\limits_{l \in \left\{ {1, \ldots ,L} \right\}} \left\{ {{p_{l,{\mathrm{in}}}^\mathrm{sc}}}\left[ n \right] \right\} \!=\!\frac{\tau_e}{\tau_c}\kappa \mathop {\max }\limits_{l \in \left\{ {1, \ldots ,L} \right\}} \left\{  {{p_{\mathrm{d}}^\mathrm{sc}}\frac{{{\Upsilon _l}\left[ n \right] \!+\! {{\left( {{\mathrm{tr}}\left( {{{\mathbf{Q}}_l}}\left[ n \right] \right) \!+\! {{\left\| {{{{\mathbf{\bar h}}}_l}}\left[ n \right] \right\|}^2}} \right)}^2}}}{{{\mathrm{tr}}\left( {{{\mathbf{Q}}_l}}\left[ n \right] \right) + {{\left\| {{{{\mathbf{\bar h}}}_l}}\left[ n \right] \right\|}^2}}}} \right\}.
\end{align}
\end{cor}
\begin{IEEEproof}
It follows the similar steps in Theorem \ref{thm1}.
\end{IEEEproof}

In SC systems, we can obtain the uplink data transmission power and the next slot pilot transmit power as $p_\text{u}^\text{sc} \left[ n \right] = \frac{\tau_c}{\tau_c-\tau_p-\tau_e}\left(1-\partial\right){P_{{\text{HE}}}^\text{sc}}\left[ n \right]$ and $p \left[ n+1 \right] = \frac{\tau_c}{\tau_p}\partial {P_{{\text{HE}}}^\text{sc}}\left[ n \right]$, respectively.

\subsubsection{Uplink SE}

In the uplink of a SC system, each AP first estimates the channels based on signals sent from the UAV, as described earlier. The so-obtained channel estimate of UAV is used to multiply the received signal for detecting the desired signal. The combined uplink signal at the AP $l$ is
\begin{align}
  &{{\mathbf{y}}_l^\text{sc}}\left[ n \right] = {\mathbf{\hat g}}_l^{\text{H}}\left[ n \right]\left( {{{\mathbf{g}}_l}\left[ n \right]\left( {\sqrt {\kappa {p_{\text{u}}^\text{sc}}\left[ n \right]} s\left[ n \right] + \eta \left[ n \right]} \right) + {{\mathbf{n}}_l}\left[ n \right]} \right) \notag \\
   &= \sqrt {\kappa {p_{\text{u}}^\text{sc}}\left[ n \right]} {\mathbf{\hat g}}_l^{\text{H}}\left[ n \right]{{{\mathbf{\hat g}}}_l}\left[ n \right]s\left[ n \right] \!+\! \underbrace {\sqrt {\kappa {p_{\text{u}}^\text{sc}}\left[ n \right]} {\mathbf{\hat g}}_l^{\text{H}}\left[ n \right]{{{\mathbf{\tilde g}}}_l}\left[ n \right]s\left[ n \right]}_{{{\mathbf{I}}_{l,1}}\left[ n \right]} \!+\! \underbrace {{\mathbf{\hat g}}_l^{\text{H}}\left[ n \right]{{\mathbf{g}}_l}\left[ n \right]\eta \left[ n \right]}_{{{\mathbf{I}}_{l,2}}\left[ n \right]} \!+\! \underbrace {{\mathbf{\hat g}}_l^{\text{H}}\left[ n \right]{{\mathbf{n}}_l}\left[ n \right]}_{{{\mathbf{I}}_{l,3}}\left[ n \right]},
\end{align}
where the first term denotes the desired received signal from UAV.
The remaining terms ${{{\mathbf{I}}_{l,1}}\left[ n \right]}$, ${{{\mathbf{I}}_{l,2}}\left[ n \right]}$ and ${{{\mathbf{I}}_{l,3}}\left[ n \right]}$ are uncorrelated and represent interference caused by channel estimation errors, the hardware impairment effect, and receiver noise at AP.
\begin{cor}
Using the maximum-ratio combining, the capacity of UAV is lower bounded as
\begin{align}
{\mathrm{SE}}^\mathrm{sc}\left[ n \right] = \frac{\tau_c-\tau_p-\tau_e}{\tau_c} \mathop {\max }\limits_{l \in \left\{ {1, \ldots ,L} \right\}} \mathbb{E}\left\{ {{{\log }_2}\left( {1 + {\mathrm{SIN}}{{\mathrm{R}}_l^\mathrm{sc}}}\left[ n \right] \right)} \right\},
\end{align}
where
\begin{align}
{\mathrm{SIN}}{{\mathrm{R}}_l^\mathrm{sc}}\left[ n \right] = \frac{{\kappa {p_{\mathrm{u}}^\mathrm{sc}}\left[ n \right]{{\left| {{\mathbf{\hat g}}_l^{\mathrm{H}}\left[ n \right]{{{\mathbf{\hat g}}}_l}\left[ n \right]} \right|}^2}}}{{\left( {1 - \kappa } \right){p_{\mathrm{u}}^\mathrm{sc}}\left[ n \right]{{\left| {{\mathbf{\hat g}}_l^{\mathrm{H}}\left[ n \right]{{{\mathbf{\hat g}}}_l}\left[ n \right]} \right|}^2} + {\mathbf{\hat g}}_l^{\mathrm{H}}\left[ n \right]\left( {{p_{\mathrm{u}}^\mathrm{sc}}\left[ n \right]{{\mathbf{C}}_l}\left[ n \right] + {\sigma ^2}{{\mathbf{I}}_N}} \right){{{\mathbf{\hat g}}}_l}\left[ n \right]}}.
\end{align}
\end{cor}
\begin{IEEEproof}
It follows similar steps in \cite[Theorem 4.1]{bjornson2017massive} for cellular massive MIMO.
\end{IEEEproof}

\subsection{Cellular massive MIMO}

We consider a cellular network with one cell and $LN$ antennas for cellular BS. It means that the UAV is always served by the cellular BS when flying from the initial position to the destination.
The block-fading channel from BS to UAV is modeled as
\begin{align}
{{\mathbf{g}}^{\text{c}}}\left[ n \right] \sim \mathcal{C}\mathcal{N}\left( {{{{\mathbf{\bar h}}}^{\text{c}}}\left[ n \right],{{\mathbf{R}}^{\text{c}}}}\left[ n \right] \right),
\end{align}
where ${{{\mathbf{\bar h}}}^{\text{c}}}\left[ n \right] \in {\mathbb{C}^{LN}}$ is the LoS component, and ${{\mathbf{R}}^{\text{c}}}\left[ n \right] \in {\mathbb{C}^{LN \times LN}}$ is the spatial correlation matrix of the NLoS components. When using standard LMMSE estimation, the LMMSE estimate of ${{\mathbf{g}}^{\text{c}}}\left[ n \right] \in {\mathbb{C}^{LN}}$ and the independent estimation error ${{{\mathbf{\tilde g}}}^{\text{c}}}\left[ n \right] \in {\mathbb{C}^{LN}}$ are respectively given by
\begin{align}
{{{\mathbf{\hat g}}}^{\text{c}}}\left[ n \right] &\sim \mathcal{C}\mathcal{N}\left( {{{{\mathbf{\bar h}}}^{\text{c}}}\left[ n \right],{{\mathbf{Q}}^{\text{c}}}\left[ n \right]} \right), \\
{{{\mathbf{\tilde g}}}^{\text{c}}}\left[ n \right] &\triangleq {{\mathbf{g}}^{\text{c}}}\left[ n \right] - {{{\mathbf{\hat g}}}^{\text{c}}}\left[ n \right] \sim \mathcal{C}\mathcal{N}\left( {{\mathbf{0}},{{\mathbf{C}}^{\text{c}}}\left[ n \right]} \right),
\end{align}
where
\begin{align}
{{\mathbf{Q}}^{\text{c}}}\left[ n \right] &= {{\mathbf{R}}^{\text{c}}}\left[ n \right] - {{\mathbf{C}}^{\text{c}}}\left[ n \right] = \kappa p\left[ n \right]{{\mathbf{R}}^{\text{c}}}\left[ n \right]{{\mathbf{\Psi }}^{\text{c}}}\left[ n \right]{{\mathbf{R}}^{\text{c}}}\left[ n \right] \\
{{\mathbf{\Psi }}^{\text{c}}}\left[ n \right] &= {\left( {p\left[ n \right]{{\mathbf{R}}^{\text{c}}}\left[ n \right] + \left( {1 - \kappa } \right)p\left[ n \right]{{{\mathbf{\bar h}}}^{\text{c}}}\left[ n \right]{{\left( {{{{\mathbf{\bar h}}}^{\text{c}}}\left[ n \right]} \right)}^{\text{H}}} + {\sigma ^2}{{\mathbf{I}}_{LN}}} \right)^{ - 1}}.
\end{align}
\subsubsection{Downlink HE}

In cellular massive MIMO system with one cell, the UAV harvests energy from the cellular BS. Then, the received signal from BS is
\begin{align}\label{r2}
r^\text{c}\left[ n \right] = \sqrt {\kappa {p_{\text{d}}^\text{c}}} \left({{\mathbf{g}}^{\text{c}}}\left[ n \right]\right)^{\text{H}} {\mathbf{w}}^\text{c}\left[ n \right]v\left[ n \right] + \mu \left[ n \right] + n\left[ n \right],
\end{align}
where $\mu\left[ n \right]  \sim \mathcal{C}\mathcal{N}\left( {0,\left( {1 - \kappa } \right){p_{{\text{in}}}^\text{c}}\left[ n \right]} \right)$ is the hardware impairment of UAV. ${p_{{\text{in}}}^\text{c}}\left[ n \right]$ is the received energy from cellular BS expressed as
\begin{align}
{p_{{\text{in}}}^\text{c}}\left[ n \right] = \mathbb{E}\left\{ {{{\left| {\sqrt {{p_{\text{d}}^\text{c}}} \left({\mathbf{g}}^{\text{c}}\left[ n \right]\right)^{\text{H}} {{\mathbf{w}}}^\text{c}\left[ n \right]} \right|}^2}} \right\}.
\end{align}
\begin{cor}
Considering the energy due to hardware impairment and the noise in \eqref{r2} cannot be harvested and using the beamformer ${\mathbf{w}}^\mathrm{c}\left[ n \right] = {{{{{\mathbf{\hat g}}}}}^\mathrm{c}\left[ n \right]}/{{\sqrt {\mathbb{E}\left\{ {{{\left\| {{{{\mathbf{\hat g}}}}^\mathrm{c}\left[ n \right]} \right\|}^2}} \right\}} }}$, the HE by UAV at time slot $n$ can be derived as
\begin{align}
{{\text{P}}_{{\mathrm{HE}}}^\mathrm{c}}\left[ n \right] = \frac{{{\tau _e}}}{{{\tau _c}}}\kappa {p_{{\mathrm{in}}}^\mathrm{c}}\left[ n \right] = \frac{\tau_e}{\tau_c}\kappa {p_{\mathrm{d}}^\mathrm{c}}\frac{{{\Upsilon}^\mathrm{c}\left[ n \right] + {{\left( {{\mathrm{tr}}\left( {\mathbf{Q}}^\mathrm{c}\left[ n \right] \right) + {{\left\| {{\mathbf{\bar h}}}^\mathrm{c}\left[ n \right] \right\|}^2}} \right)}^2}}}{{{\mathrm{tr}}\left( {\mathbf{Q}}^\mathrm{c}\left[ n \right] \right) + {{\left\| {{\mathbf{\bar h}}}^\mathrm{c}\left[ n \right] \right\|}^2}}},
\end{align}
where
\begin{align}
{\Upsilon^\mathrm{c}}\left[ n \right] = {\mathrm{tr}}\left( {{{\mathbf{R}}}^\mathrm{c}\left[ n \right]{{\mathbf{Q}}}^\mathrm{c}\left[ n \right]} \right) + \left({\mathbf{\bar h}}^{\mathrm{c}}\left[ n \right]\right)^{\mathrm{H}} {{\mathbf{Q}}}^\mathrm{c}\left[ n \right]{{{\mathbf{\bar h}}}}^\mathrm{c}\left[ n \right] + \left({\mathbf{\bar h}}^{\mathrm{c}}\left[ n \right]\right)^{\mathrm{H}} {{\mathbf{R}}}^\mathrm{c}\left[ n \right]{{{\mathbf{\bar h}}}}^\mathrm{c}\left[ n \right].
\end{align}
\end{cor}
\begin{IEEEproof}
It follows the similar steps in Theorem \ref{thm1}.
\end{IEEEproof}

In cellular systems, we can obtain the uplink data transmission power and the next slot pilot transmit power as $p_\text{u}^\text{c} \left[ n \right] = \frac{\tau_c}{\tau_c-\tau_p-\tau_e}\left(1-\partial\right){P_{{\text{HE}}}^\text{c}}\left[ n \right]$ and $p \left[ n+1 \right] = \frac{\tau_c}{\tau_p}\partial {P_{{\text{HE}}}^\text{c}}\left[ n \right]$, respectively.

\subsubsection{Uplink SE}

The received signal at BS is
\begin{align}
  {{\mathbf{y}}^{\text{c}}} &= {\left( {{{{\mathbf{\hat g}}}^{\text{c}}}\left[ n \right]} \right)^{\text{H}}}\left( {{{\mathbf{g}}^{\text{c}}}\left[ n \right]\left( {\sqrt {\kappa {p_{\text{u}}^\text{c}}\left[ n \right]} s\left[ n \right] + \eta \left[ n \right]} \right) + {{\mathbf{n}}_l}\left[ n \right]} \right) = \sqrt {\kappa {p_{\text{u}}^\text{c}}\left[ n \right]} {\left( {{{{\mathbf{\hat g}}}^{\text{c}}}\left[ n \right]} \right)^{\text{H}}}{{{\mathbf{\hat g}}}^{\text{c}}}\left[ n \right]s\left[ n \right] \notag\\
   &+ \underbrace {\sqrt {\kappa {p_{\text{u}}^\text{c}}\left[ n \right]} {{\left( {{{{\mathbf{\hat g}}}^{\text{c}}}\left[ n \right]} \right)}^{\text{H}}}{{{\mathbf{\tilde g}}}^{\text{c}}}\left[ n \right]s\left[ n \right]}_{{{\mathbf{\Gamma }}_1}\left[ n \right]} + \underbrace {{{\left( {{{{\mathbf{\hat g}}}^{\text{c}}}\left[ n \right]} \right)}^{\text{H}}}{{\mathbf{g}}^{\text{c}}}\left[ n \right]\eta \left[ n \right]}_{{{\mathbf{\Gamma }}_2}\left[ n \right]} + \underbrace {{{\left( {{{{\mathbf{\hat g}}}^{\text{c}}}\left[ n \right]} \right)}^{\text{H}}}{{\mathbf{n}}_l}\left[ n \right]}_{{{\mathbf{\Gamma }}_3}\left[ n \right]},
\end{align}
where the first term denotes the desired received signal from UAV.
The remaining terms ${{{\mathbf{\Gamma }}_1}\left[ n \right]}$, ${{{\mathbf{\Gamma }}_2}\left[ n \right]}$ and ${{{\mathbf{\Gamma }}_3}\left[ n \right]}$ are uncorrelated and represent interference caused by channel estimation errors, the hardware impairment effect, and receiver noise at AP.
\begin{cor}
Using the maximum-ratio combining, the capacity of UAV is lower bounded as
\begin{align}
{\mathrm{SE}}^\mathrm{c}\left[ n \right] = \frac{\tau_c-\tau_p-\tau_e}{\tau_c} \mathbb{E}\left\{ {{{\log }_2}\left( {1 +  \mathrm{SINR}^\mathrm{c}\left[ n \right] } \right)} \right\},
\end{align}
where $\mathrm{SINR}^\mathrm{c}\left[ n \right]$ is given by
\begin{align}
{\mathrm{SINR}}^\mathrm{c}\left[ n \right] = \frac{{\kappa {p_{\mathrm{u}}^\mathrm{c}}\left[ n \right]{{\left| {{{\left( {{{{\mathbf{\hat g}}}^{\mathrm{c}}}\left[ n \right]} \right)}^{\mathrm{H}}}{{{\mathbf{\hat g}}}^{\mathrm{c}}}\left[ n \right]} \right|}^2}}}{{\left( {1 - \kappa } \right){p_{\mathrm{u}}^\mathrm{c}}\left[ n \right]{{\left| {{{\left( {{{{\mathbf{\hat g}}}^{\mathrm{c}}}\left[ n \right]} \right)}^{\mathrm{H}}}{{{\mathbf{\hat g}}}^{\mathrm{c}}}\left[ n \right]} \right|}^2} + {{\left( {{{{\mathbf{\hat g}}}^{\mathrm{c}}}\left[ n \right]} \right)}^{\mathrm{H}}}\left( {{p_{\mathrm{u}}^\mathrm{c}}\left[ n \right]{\mathbf{C}}^\mathrm{c}\left[ n \right] + {\sigma ^2}{{\mathbf{I}}_N}} \right){{{\mathbf{\hat g}}}^{\mathrm{c}}}\left[ n \right]}}.
\end{align}
\end{cor}
\begin{IEEEproof}
It follows similar steps in \cite[Theorem 4.1]{bjornson2017massive} for cellular massive MIMO.
\end{IEEEproof}

\subsection{Interference From One Terrestrial UE}\label{interference}

The terrestrial interference has a significant impact on UAV communications \cite{zeng2019accessing}. In the following, we assume that there is a single-antenna TUE in considered systems. The TUE and UAV respectively use one channel use for channel estimation. In addition, there is no hardware impairment and HE at TUE, and TUE has the same length of uplink and downlink phases with UAV for data transmission.
Let ${{\mathbf{h}}_{{\text{te}},l}}$ denote the channel coefficient between the TUE and the $l$th AP. The fading channel ${{\mathbf{h}}_{{\text{te}},l}}$ is modelled as
\begin{align}
{{\mathbf{h}}_{{\text{te}},l}} \sim \mathcal{C}\mathcal{N}\left( {{\mathbf{0}},{{\mathbf{R}}_{{\text{te}},l}}} \right),l = 1, \ldots ,L,
\end{align}
where ${{{\mathbf{R}}_{{\text{te}},l}}} \in {\mathbb{C}^{N \times N}}$ is the spatial correlation matrix and ${\beta _{{\text{te}},l}} = {\text{tr}}\left( {{{\mathbf{R}}_{{\text{te}},l}}} \right)/N$ is the large-scale fading coefficient.
Using minimum mean square error (MMSE) estimation, the channel estimation ${{{\mathbf{\hat h}}}_{{\text{te}},l}}$ can be obtained as
\begin{align}
{{{\mathbf{\hat h}}}_{{\text{te}},l}} = \sqrt {{p_{{\text{te}}}}} {{\mathbf{R}}_{{\text{te}},l}}{{\mathbf{\Psi }}_{{\text{te}},l}}{{\mathbf{z}}_l},
\end{align}
where ${{p_{{\text{te}}}}}$ is the signal transmit power of TUE, ${{\mathbf{\Psi }}_{{\text{te}},l}} = {\left( {{p_{{\text{te}}}}{{\mathbf{R}}_{{\text{te}},l}} + {\sigma ^2}{{\mathbf{I}}_N}} \right)^{ - 1}}$, and ${{\mathbf{z}}_l} = {{\mathbf{h}}_{{\text{te}},l}}\sqrt {{p_{{\text{te}}}}}  + {{\mathbf{n}}_l}$ is the received signal between AP $l$ and TUE.
The channel estimation ${{{\mathbf{\hat h}}}_{{\text{te}},l}}$ is distributed as $\mathcal{C}\mathcal{N}\left( {{\mathbf{0}},{{\mathbf{G}}_{{\text{te}},l}}} \right)$, where
\begin{align}
{{\mathbf{G}}_{{\text{te}},l}} = {p_{{\text{te}}}}{{\mathbf{R}}_{{\text{te}},l}}{{\mathbf{\Psi }}_{{\text{te}},l}}{{\mathbf{R}}_{{\text{te}},l}}.
\end{align}
\subsubsection{Downlink HE}
For CF massive MIMO systems with maximum ratio precoding to TUE, the transmitted signal from AP $l$ at time slot $n$ can be expressed as
\begin{align}
{\mathbf{x}}_l^{{\text{cf}}}[n] = \sqrt {p_{\text{d}}^{{\text{cf}}}} {\mathbf{w}}[n]{v_l}[n] + \sqrt {p_{\text{d}}^{{\text{cf}}}} {{{\mathbf{\hat h}}}_{{\text{te}},l}}[n]{x_l}[n].
\end{align}
Then, the received power in UAV at time slot $n$ can be expressed as
\begin{align}
p_{{\text{in }}}^{{\text{cf }}}[n] = \sum\limits_{l = 1}^L \mathbb{E} \left\{ {{{\left| {\sqrt {p_{\text{d}}^{{\text{cf}}}} {\mathbf{g}}_l^{\text{H}}[n]{\mathbf{w}}[n]} \right|}^2}} \right\} + \sum\limits_{l = 1}^L {\mathbb{E}\left\{ {{{\left| {\sqrt {p_{\text{d}}^{{\text{cf}}}} {\mathbf{g}}_l^{\text{H}}[n]{{{\mathbf{\hat h}}}_{{\text{te}},l}}[n]} \right|}^2}} \right\}}.
\end{align}
Following similar steps in Theorem \ref{thm1}, we can obtain the downlink HE ${P_{{\mathrm{HE}}}^\mathrm{cf}}\left[ n \right] $ as
\begin{align}
\frac{\tau_e}{\tau_c} \kappa   {p_{\mathrm{d}}^\mathrm{cf}}\sum\limits_{l = 1}^L \!\!\left(\! {\frac{{ {\Upsilon _l}\left[ n \right]  \!+\! {\left( {\mathrm{tr}}\left( {{{\mathbf{Q}}_l}}\left[ n \right] \right) \!+\! {{\left\| {{{{\mathbf{\bar h}}}_l}}\left[ n \right] \right\|}^2} \right)^2} }}{{{\mathrm{tr}}\left( {{{\mathbf{Q}}_l}}\left[ n \right] \right) + {{\left\| {{{{\mathbf{\bar h}}}_l}}\left[ n \right] \right\|}^2}}} \!+\! {{\text{tr}}\left( {{{\mathbf{G}}_{{\text{te}},l}\left[ n \right]}{{\mathbf{R}}_l}\left[ n \right]} \right) \!+\! {\mathbf{\bar h}}_l^{\text{H}}\left[ n \right]{{\mathbf{G}}_{{\text{te}},l}\left[ n \right]}{{{\mathbf{\bar h}}}_l\left[ n \right]}}}   \!\right).
\end{align}
For SC systems, we assume that the TUE is served by AP ${l'}$. Then, we can write ${P_{{\mathrm{HE}}}^\mathrm{sc}}\left[ n \right] $ as
\begin{align}
\!\!\!\frac{\tau_e}{\tau_c}\kappa p_{\text{d}}^{{\text{sc}}}\! \mathop {\max }\limits_{l \in \left\{ {1, \ldots ,L} \right\}} \!\!\left\{\!\!  {\frac{{{\Upsilon _l}\!\left[ n \right] \!+\! {{\left(\! {{\mathrm{tr}}\left( {{{\mathbf{Q}}_l}}\!\left[ n \right] \right) \!+\! {{\left\| {{{{\mathbf{\bar h}}}_l}}\!\left[ n \right] \right\|}^2}} \!\right)}^2}}}{{{\mathrm{tr}}\left( {{{\mathbf{Q}}_l}}\!\left[ n \right] \right) + {{\left\| {{{{\mathbf{\bar h}}}_l}}\left[ n \right] \right\|}^2}}}} \!+\! { {{\text{tr}}\left( {{{\mathbf{G}}_{{\text{te}},l'}\!\left[ n \right]}{{\mathbf{R}}_{l'}\!\left[ n \right]}} \right) \!+\! {\mathbf{\bar h}}_{l'}^{\text{H}}\!\left[ n \right]{{\mathbf{G}}_{{\text{te}},l'}\!\left[ n \right]}{{{\mathbf{\bar h}}}_{l'}\!\left[ n \right]}} } \!\!\right\}\!.
\end{align}
For cellular massive MIMO systems, we can write ${P_{{\mathrm{HE}}}^\mathrm{c}}\left[ n \right] $ as
\begin{align}
\frac{\tau_e}{\tau_c}\kappa {p_{\mathrm{d}}^\mathrm{c}}\left(\frac{{{\Upsilon}^\mathrm{c}\left[ n \right] \!+\! {{\left( {{\mathrm{tr}}\left( {\mathbf{Q}}^\mathrm{c}\left[ n \right] \right) \!+\! {{\left\| {{\mathbf{\bar h}}}^\mathrm{c}\left[ n \right] \right\|}^2}} \right)}^2}}}{{{\mathrm{tr}}\left( {\mathbf{Q}}^\mathrm{c}\left[ n \right] \right) + {{\left\| {{\mathbf{\bar h}}}^\mathrm{c}\left[ n \right] \right\|}^2}}} \!+\! {{\text{tr}}\left( {{\mathbf{G}}_{{\text{te}}}^{\text{c}}\left[ n \right]{{\mathbf{R}}^{\text{c}}\left[ n \right]}} \right) \!+\! {{\left( {{{{\mathbf{\bar h}}}^{\text{c}}\left[ n \right]}} \right)}^{\text{H}}}{\mathbf{G}}_{{\text{te}}}^{\text{c}}\left[ n \right]{{{\mathbf{\bar h}}}^{\text{c}}\left[ n \right]}} \right).
\end{align}

\subsubsection{Uplink SE}

For CF massive MIMO systems with the TUE, the received complex baseband signal in AP $l$ at time slot $n$ can be expressed as
\begin{align}
{\mathbf{y}}_l^{{\text{cf}}}[n] = {{\mathbf{g}}_l}[n]\left( {\sqrt {\kappa p_{\text{u}}^{{\text{cf}}}[n]} s[n] + \eta [n]} \right) + {{\mathbf{h}}_{{\text{te}},l}}\left[ n \right]\sqrt {p_{{\text{te,u}}}^{{\text{cf}}}} x[n] + {{\mathbf{n}}_l}[n],
\end{align}
where ${p_{{\text{te,u}}}^{{\text{cf}}}}$ is the uplink data transmission power of the TUE. Based on \eqref{CPU}, we can obtain the received data at CPU as
\begin{align}
\hat s\left[ n \right] = {\text{DS}}\left[ n \right] + {\text{HI}}\left[ n \right] + {\text{BU}}\left[ n \right] + \underbrace {\sum\limits_{l = 1}^L {{\mathbf{\hat g}}_l^{\text{H}}[n]{{\mathbf{h}}_{{\text{te}},l}}\sqrt {p_{{\text{te,u}}}^{{\text{cf}}}} x[n]} }_{{\text{UI}}\left[ n \right]} + {\text{NS}}\left[ n \right],
\end{align}
where ${{\text{UI}}\left[ n \right]}$ represents the interference caused by transmitted
data from TUE. Following similar steps in Theorem \ref{thm2}, the capacity of UAV is lower bounded by
\begin{align}
{\text{SE}}\left[ n \right] =\frac{1\!-\!\tau_p\!-\!\tau_e}{\tau_c} \log \!\left(\! {1 \!+\! \frac{{\mathbb{E}\!\left\{ {{{\left| {{\text{DS}}}\left[ n \right] \right|}^2}} \right\}}}{{\mathbb{E}\!\left\{ {{{\left| {{\text{BU}}}\left[ n \right] \right|}^2}} \right\} \!+\! \mathbb{E}\!\left\{ {{{\left| {{\text{HI}}}\left[ n \right] \right|}^2}} \right\} \!+\! \mathbb{E}\!\left\{ {{{\left| {{\text{UI}}}\left[ n \right] \right|}^2}} \right\} \!+\! \mathbb{E}\!\left\{ {{{\left| {{\text{NS}}}\left[ n \right] \right|}^2}} \right\}}}} \!\right),
\end{align}
where
\begin{align}
\mathbb{E}\left\{ {{{\left| {{\text{UI}}\left[ n \right]} \right|}^2}} \right\} = \mathbb{E}\left\{ {{{\left| {\sum\limits_{l = 1}^L {{\mathbf{\hat g}}_l^{\text{H}}[n]{{\mathbf{h}}_{{\text{te}},l}}\sqrt {p_{{\text{te,u}}}^{{\text{cf}}}} } } \right|}^2}} \right\} = p_{{\text{te,u}}}^{{\text{cf}}}\sum\limits_{l = 1}^L \left( {{\text{tr}}\left( {{{\mathbf{R}}_{{\text{te}},l}}{{\mathbf{Q}}_l}} \right) + {\mathbf{\bar h}}_l^{\text{H}}{{\mathbf{R}}_{{\text{te}},l}}{{{\mathbf{\bar h}}}_l}} \right) .
\end{align}
\begin{cor}
With the help of LSFD receiver cooperation \cite{bjornson2019making}, when there is a TUE as interference, we can derive the maximum SE as
\begin{align}
{\mathrm{S}}{{\mathrm{E}}^{{\mathrm{LSFD}}}}[n] = \frac{{1 - {\tau _p} - {\tau _e}}}{{{\tau _c}}}\log \left( {1 + {\mathrm{SIN}}{{\mathrm{R}}^{{\mathrm{LSFD}}}}\left[ n \right]} \right),
\end{align}
with ${\mathrm{SIN}}{{\mathrm{R}}^{{\mathrm{LSFD}}}}\left[ n \right]$ is given by
\begin{align}
\kappa p_{\mathrm{u}}^{{\mathrm{cf}}}[n]{{\mathbf{b}}^{\mathrm{H}}}\left[ n \right]{\left( {p_{\mathrm{u}}^{{\mathrm{cf}}}[n]{\mathbf{\Gamma }}\left[ n \right] + (1 - \kappa )p_{\mathrm{u}}^{{\mathrm{cf}}}[n]{\mathbf{b}}\left[ n \right]{{\mathbf{b}}^{\mathrm{H}}}\left[ n \right] + p_{{\mathrm{te,u}}}^{{\mathrm{cf}}}{\mathbf{{\mathbf{T}}}}\left[ n \right] + {\sigma ^2}{\mathbf{\Lambda }}\left[ n \right]} \right)^{ - 1}}{\mathbf{b}}\left[ n \right],
\end{align}
where
\begin{align}
{\mathbf{b}}\left[ n \right] &= {\left[ {\left( {\mathrm{tr} \left( {{{\mathbf{Q}}_1}[n]} \right) + {{\left\| {{{\overline {\mathbf{h}} }_1}[n]} \right\|}^2}} \right) \ldots \left( \mathrm{tr} \left( {{{\mathbf{Q}}_L}[n]} \right) + {{\left\| {{{\overline {\mathbf{h}} }_L}[n]} \right\|}^2}\right) } \right]^{\mathrm{T}}}, \notag\\
{\mathbf{\Gamma }}\left[ n \right] &= {\mathrm{diag}}\left( {{\Upsilon _1}\left[ n \right], \ldots ,{\Upsilon _L}\left[ n \right]} \right), \notag\\
{\mathbf{{\mathbf T}}}\left[ n \right] &= {\mathrm{diag}}\left( {\left( {{\mathrm{tr}}\left( {{{\mathbf{R}}_{{\mathrm{te}},1}}{{\mathbf{Q}}_1}} \right) + {\mathbf{\bar h}}_1^{\mathrm{H}}{{\mathbf{R}}_{{\mathrm{te}},1}}{{{\mathbf{\bar h}}}_1}} \right), \ldots ,\left( {{\mathrm{tr}}\left( {{{\mathbf{R}}_{{\mathrm{te}},L}}{{\mathbf{Q}}_L}} \right) + {\mathbf{\bar h}}_L^{\mathrm{H}}{{\mathbf{R}}_{{\mathrm{te}},L}}{{{\mathbf{\bar h}}}_L}} \right)} \right), \notag\\
{\mathbf{\Lambda }}\left[ n \right] &= {\mathrm{diag}}\left( {\left( {\mathrm{tr} \left( {{{\mathbf{Q}}_1}[n]} \right) + {{\left\| {{{\overline {\mathbf{h}} }_1}[n]} \right\|}^2}} \right), \ldots ,\left( {\mathrm{tr} \left( {{{\mathbf{Q}}_L}[n]} \right) + {{\left\| {{{\overline {\mathbf{h}} }_L}[n]} \right\|}^2}} \right)} \right).
\end{align}
\end{cor}
\begin{IEEEproof}
It follows similar steps in \cite[Corollary 2]{bjornson2019making} for CF massive MIMO with LSFD.
\end{IEEEproof}
For SC systems, a lower bound on the capacity of UAV is
\begin{align}
{\mathrm{SE}}^\mathrm{sc}\left[ n \right] = \frac{\tau_c-\tau_p-\tau_e}{\tau_c} \mathop {\max }\limits_{l \in \left\{ {1, \ldots ,L} \right\}} \mathbb{E}\left\{ {{{\log }_2}\left( {1 + {\mathrm{SIN}}{{\mathrm{R}}_l^\mathrm{sc}}}\left[ n \right] \right)} \right\},
\end{align}
where ${\mathrm{SIN}}{{\mathrm{R}}_l^\mathrm{sc}}\left[ n \right]$ is given by
\begin{align}
\frac{{\kappa {p_{\mathrm{u}}^\mathrm{sc}}\left[ n \right]{{\left| {{\mathbf{\hat g}}_l^{\mathrm{H}}\left[ n \right]{{{\mathbf{\hat g}}}_l}\left[ n \right]} \right|}^2}}}{{\left( {1 - \kappa } \right){p_{\mathrm{u}}^\mathrm{sc}}\left[ n \right]{{\left| {{\mathbf{\hat g}}_l^{\mathrm{H}}\left[ n \right]{{{\mathbf{\hat g}}}_l}\left[ n \right]} \right|}^2} + {\mathbf{\hat g}}_l^{\mathrm{H}}\left[ n \right]\left( {{p_{\mathrm{u}}^\mathrm{sc}}\left[ n \right]{{\mathbf{C}}_l}\left[ n \right] + {p_{{\text{te,u}}}^{{\text{sc}}}\left[ n \right]{{\mathbf{R}}_{{\text{te}},l}}\left[ n \right]} + {\sigma ^2}{{\mathbf{I}}_N}} \right){{{\mathbf{\hat g}}}_l}\left[ n \right]}}.
\end{align}
For cellular MIMO systems, the capacity of UAV is lower bounded by
\begin{align}
{\mathrm{SE}}^\mathrm{c}\left[ n \right] = \frac{\tau_c-\tau_p-\tau_e}{\tau_c} \mathbb{E}\left\{ {{{\log }_2}\left( {1 +  \mathrm{SINR}^\mathrm{c}\left[ n \right] } \right)} \right\},
\end{align}
where $\mathrm{SINR}^\mathrm{c}\left[ n \right]$ is given by
\begin{align}
\frac{{\kappa {p_{\mathrm{u}}^\mathrm{c}}\left[ n \right]{{\left| {{{\left( {{{{\mathbf{\hat g}}}^{\mathrm{c}}}\left[ n \right]} \right)}^{\mathrm{H}}}{{{\mathbf{\hat g}}}^{\mathrm{c}}}\left[ n \right]} \right|}^2}}}{{\left( {1 - \kappa } \right){p_{\mathrm{u}}^\mathrm{c}}\left[ n \right]{{\left| {{{\left( {{{{\mathbf{\hat g}}}^{\mathrm{c}}}\left[ n \right]} \right)}^{\mathrm{H}}}{{{\mathbf{\hat g}}}^{\mathrm{c}}}\left[ n \right]} \right|}^2} + {{\left( {{{{\mathbf{\hat g}}}^{\mathrm{c}}}\left[ n \right]} \right)}^{\mathrm{H}}}\left( {{p_{\mathrm{u}}^\mathrm{c}}\left[ n \right]{\mathbf{C}}^\mathrm{c}\left[ n \right] + {p_{{\text{te,u}}}^{\text{c}}\left[ n \right]{\mathbf{R}}_{{\text{te}}}^{\text{c}}\left[ n \right]} + {\sigma ^2}{{\mathbf{I}}_N}} \right){{{\mathbf{\hat g}}}^{\mathrm{c}}}\left[ n \right]}}.
\end{align}

\section{Trajectory Design}\label{se:trajectory}

In this section, the UAV trajectory designs are investigated to further improve the system performance. Meanwhile, the system performance of the proposed schemes is evaluated via per-slot SE when the UAV flies to the destination. Then, we formulate designing the trajectory of UAV to maximize the per-slot SE. Considering the velocity constraint of UAV, optimization problem can be modeled as
\begin{align}
  \mathop {\max }\limits_{{{\mathbf{q}}_{\text{u}}}\left[ n \right]} \;\;&{\text{SE}}\left[ n \right] \hfill \\
  s.t.\;\;&\left\| {{{\mathbf{q}}_{\text{u}}}\left[ n \right] - {{\mathbf{q}}_{\text{u}}}\left[ {n - 1} \right]} \right\| = {d_{\min }} = {V_\text{hor}}T , \\
   &{{\mathbf{q}}_{{\text{des}}}} = \left[ {{x_{{\text{des}}}},{y_{{\text{des}}}}} \right].
\end{align}
where $V_\text{hor}$ is the constant velocity of UAV, which places restrictions on the mobility of UAV. $T$ is the length of one coherence block and $d_\text{min}$ is the minimum flight distance for UAV in one coherence block. However, due to the optimization problem is non-convex, global optimum trajectory is hard to find.
For maximizing the per-slot SE, we thus propose effective angle search trajectory design scheme based the tabu search algorithm. In addition, AP search trajectory scheme is proposed to reduce the number of UAV direction switching. Then, line path trajectory scheme is used for comparison.

\subsection{Angle search trajectory design scheme}

Tabu search algorithm is a meta-heuristic algorithm, which explores the solution space by moving from a feasible solution to the best solution in a subset of its neighborhood with each iteration \cite{srinidhi2011layered}. Tabu search algorithm starts with an initial solution and select a best vector among the neighboring vectors even though the best current vector is worse. In addition, the tabu list is provided to makes sure high-quality search among the
solution vectors. Although these allows the algorithm to escape from the local optimum value, it also lead to UAV fly a long time near the best point. This will waste a lot of time when the UAV flies to the designated target. Based on the tabu search algorithm, we propose angle search trajectory scheme by limiting the search area of UAV flight direction, which is practical in UAV trajectory to improve the per-slot SE. The main steps of angle search trajectory scheme is presented as follows:

\emph{Step 1:} At first, we calculate the initial SE at the $0$th slot based on the given initial UAV position vector ${{\mathbf{q}}_{\text{u}}}\left[ 0 \right] = \left[ {{x_{\text{u}}}\left[ 0 \right],{y_{\text{u}}}\left[ 0 \right]} \right]$.

\emph{Step 2:} Let the last UAV position ${{\mathbf{q}}_{\text{u}}}\left[ n-1 \right]$ as the center and the minimum flight distance $d_\text{min}$ as the radius, selecting $M$ neighbouring position vectors $\left\{ {{\mathbf{q}}_{\text{u}}^1\left[ n \right], \ldots ,{\mathbf{q}}_{\text{u}}^M\left[ n \right]} \right\}$ with equiangular interval in the quadrant of the designated target ${\mathbf{q}}_{\text{des}}$. Then, we select the SE-maximizing position vector ${{\mathbf{q}}_{\text{u}}}\left[ n \right] = \max \left\{ {{\mathbf{q}}_{\text{u}}^1\left[ n \right], \ldots ,{\mathbf{q}}_{\text{u}}^M\left[ n \right]} \right\}$ as the UAV position at the $n$th slot.

\emph{Step 3:} Save the position and SE at the $n$th slot in the UAV trajectory list and per-slot SE list, respectively. Then, go to step 2 until the UAV arrives the designated target ${{\mathbf{q}}_{{\text{des}}}}$ or the maximum number of slots $N_\text{slot}$ is reached.

The aforementioned angle search trajectory design scheme is briefly given in Algorithm \ref{angle_search}.

\begin{algorithm}[htb]
\caption{Angle search trajectory design scheme.}
\label{angle_search}
\begin{algorithmic}[1]
\Require
$L, N, \kappa, H, \rho, N_\text{set}, N_\text{slot}$, initial position ${{\mathbf{q}}_{\text{u}}}\left[ 0 \right]$;
\Ensure
per-slot positions ${{\mathbf{q}}_{\text{u}}}\left[ n \right]$, per-slot ${\text{SE}}\left[ n \right]$;
\For{$r=1$ to $N_\text{set}$}
\State Randomly generate the positions of AP within a square of size $100{\text{m}} \times 100{\text{m}}$;
\State Based on the initial ${{\mathbf{q}}_{\text{u}}}\left[ 0 \right]$ to calculate the $\text{SE}\left[ 0 \right]$ as the initial historical result;
\For{$n=1$ to $N_\text{slot}$}
\State Based on ${{\mathbf{q}}_{\text{u}}}\left[ n-1 \right]$ as the center and $d_\text{min}$ as the radius, determine neighborhood;
\State Take the quadrant of ${{\mathbf{q}}_{{\text{des}}}}$ as the search direction;
\State Select $M$ neighbouring position vectors $\left\{ {{\mathbf{q}}_{\text{u}}^1\left[ n \right], \ldots ,{\mathbf{q}}_{\text{u}}^M\left[ n \right]} \right\}$;
\State Calculate the corresponding $\left\{ {{\text{S}}{{\text{E}}^1}\left[ n \right], \ldots ,{\text{S}}{{\text{E}}^M}\left[ n \right]} \right\}$;
\State According to the SE to sort all the neighbouring positions in descending order;
\State Save the best SE and the SE-maximizing position in the trajectory and SE list;
\If{UAV arrive the destination}
\State \textbf{break};
\EndIf
\EndFor
\EndFor
\State \Return {the UAV trajectory and per-slot SE list ${{\mathbf{q}}_{\text{u}}}\left[ n \right]$, ${\text{SE}}\left[ n \right]$.}
\end{algorithmic}
\end{algorithm}

\subsection{AP search trajectory design scheme}

In order to reduce the number of UAV direction switching, we provide another meta-heuristic algorithm named as AP search trajectory design scheme. From the initial position ${{\mathbf{q}}_{\text{u}}}\left[ 0 \right]$, the UAV selects the nearest AP in the quadrant of the designated target ${\mathbf{q}}_{\text{des}}$ as the direction of flight. When the UAV arrives the nearest AP, the next nearest AP in the quadrant of the designated target is taken as the new direction of flight until the UAV arrives the designated target or the maximum number of slots is reached. The aforementioned AP search trajectory design scheme is briefly given in Algorithm \ref{AP_search}.

\begin{algorithm}[htb]
\caption{AP search trajectory design scheme.}
\label{AP_search}
\begin{algorithmic}[1]
\Require
$L, N, \kappa, H, \rho, N_\text{set}, N_\text{slot}$, initial position ${{\mathbf{q}}_{\text{u}}}\left[ 0 \right]$;
\Ensure
per-slot positions ${{\mathbf{q}}_{\text{u}}}\left[ n \right]$, per-slot ${\text{SE}}\left[ n \right]$;
\For{$r=1$ to $N_\text{set}$}
\State Randomly generate the positions of AP within a square of size $100{\text{m}} \times 100{\text{m}}$;
\State Based on the initial ${{\mathbf{q}}_{\text{u}}}\left[ 0 \right]$ to calculate the $\text{SE}\left[ 0 \right]$ as the initial historical result;
\State Define ${{\mathbf{q}}_{{\text{des}}}}$ and all APs as direction nodes;
\State In the quadrant of ${{\mathbf{q}}_{{\text{des}}}}$, find the nearest direction node to ${{\mathbf{q}}_{\text{u}}}\left[ 0 \right]$;
\State Select the direction of the nearest direction node as the flight direction;
\For{$n=1$ to $N_\text{slot}$}
\If{UAV arrive the nearest direction node}
\State In the quadrant of ${{\mathbf{q}}_{{\text{des}}}}$, find the new nearest direction nodes to ${{\mathbf{q}}_{\text{u}}}\left[ n-1 \right]$;
\State Select the direction of the new nearest direction node as the new flight direction;
\EndIf
\State Take ${{\mathbf{q}}_{\text{u}}}\left[ n-1 \right]$ as the initial position, $d_\text{min}$ as the flight interval, and fly to ${{\mathbf{q}}_{\text{u}}}\left[ n \right]$;
\State Calculate the corresponding ${\text{SE}}\left[ n \right]$;
\State Save ${\text{SE}}\left[ n \right]$ and ${{\mathbf{q}}_{\text{u}}}\left[ n \right]$ in the trajectory and SE list;
\If{UAV arrive the destination}
\State \textbf{break};
\EndIf
\EndFor
\EndFor
\State \Return {the UAV trajectory and per-slot SE list ${{\mathbf{q}}_{\text{u}}}\left[ n \right]$, ${\text{SE}}\left[ n \right]$.}
\end{algorithmic}
\end{algorithm}

\subsection{Line path and all APs path trajectory design scheme}

The line path trajectory design scheme is considered for comparison. From the initial position ${{\mathbf{q}}_{\text{u}}}\left[ 0 \right]$, the UAV selects the designated target ${{\mathbf{q}}_{{\text{des}}}}$ as the direction of flight until the UAV arrives the designated target or the maximum number of slots is reached.
In addition, if the time cost of arriving at the destination is not considered, the UAV can travel all APs for comparison.
\subsection{Complexity analysis}

Because the hovering of UAV is not considered, the path length of these four flight schemes are different. We assume that the flight duration of these four flight schemes are $N_{{\text{slot}}}^{{\text{ang}}}$, $N_{{\text{slot}}}^{{\text{ap}}}$, $N_{{\text{slot}}}^{{\text{line}}}$ and $N_{{\text{slot}}}^{{\text{all}}}$, respectively.
It is not hard to see that $N_{{\text{slot}}}^{{\text{ang}}} > N_{{\text{slot}}}^{{\text{line}}}$, $N_{{\text{slot}}}^{{\text{ap}}} > N_{{\text{slot}}}^{{\text{line}}}$ and $N_{{\text{slot}}}^{{\text{all}}} \gg N_{{\text{slot}}}^{{\text{ang}}}$. In addition, the UAV must fly over some APs in AP search trajectory design scheme. We define the number of APs directly below AP search UAV trajectory as $N_\text{ap}$. Moreover, direction switching times and direction search times of UAV can indicate the complexity the trajectory design schemes.
We compare the difference among angle search, AP search, line path and all APs path trajectory design schemes as in TABLE \ref{table}.
Due to ${N^{{\text{ap}}}} \ll N_{{\text{slot}}}^{{\text{ang}}}$, we can come to a conclusion that angle search trajectory design scheme has the highest complexity, and the complexity of line path trajectory design scheme is minimum.

\begin{table}[t]
\caption{Complexity comparison among angle search, AP search, line path and all APs path.}
\vspace{4mm}
\centering
\setlength{\tabcolsep}{5mm}{
\begin{tabular}{|c|c|c|c|c|}
    \hline
    \hline
    Different trajectory design schemes & Angle search & AP search & Line path & All APs path \cr\hline
    Flight duration  & $N_{{\text{slot}}}^{{\text{ang}}}$ & $N_{{\text{slot}}}^{{\text{ap}}}$ & $N_{{\text{slot}}}^{{\text{line}}}$ & $N_{{\text{slot}}}^{{\text{all}}} \gg N_{{\text{slot}}}^{{\text{ang}}}$ \cr\hline
    APs directly below UAV trajectory  & ----- & $ {N^{{\text{ap}}}} $ & ----- & $L$ \cr\hline
    Direction switching times & $<N_{{\text{slot}}}^{{\text{ang}}}$ & ${N^{{\text{ap}}}}+1$ & 1 & $L+1$ \cr\hline
    Direction search times & $M N_{{\text{slot}}}^{{\text{ang}}}$ & $<L\left({N^{{\text{ap}}}}+1\right)$ & 1 & $L+1$ \cr\hline
    \hline
\end{tabular}}
\label{table}
\end{table}

\section{Numerical Results and Discussion}\label{se:numerical}

Based on the simulation setup in \cite{shaik2020mmse}, we consider a simulation setup where $L=20$ APs and one UAV are independently and randomly distributed within a square of size $100\text{m}\times100\text{m}$.
The carrier frequency is set as $f_c=2$ GHz, the initial pilot transmit power $p\left[ 0 \right]=1$ dBm, and the signal transmit power of TUE $p_\text{te} = p_\text{te,u} = 1$ dBm. The bandwidth is $B=20$ MHz, the reference channel power gain $\beta_0=-40$ dB \cite{lyu2018uav}, and the noise power is $\sigma^2=-96$ dBm. The constant velocity of UAV is $V_\text{hor}=20$ m/s, and the length of coherence block is $T=2$ ms\footnote{The coherence block time of 2 ms in this scenario can support high mobility of 20 m/s in practice \cite{bjornson2017massive}.} including $\tau_c=200$ channel users \cite{bjornson2019making}. To make a fair comparison among CF massive MIMO, SC and cellular massive MIMO systems, we set the downlink transmit power as $p_{\text{d}}^{{\text{sc}}}\left[ n \right] = Lp_{\text{d}}^{{\text{cf}}}\left[ n \right]$, $p_{\text{d}}^{{\text{cf}}}\left[ n \right] = p_{\text{d}}^{{\text{c}}}\left[ n \right] = 30$ dBm \cite{Ngo2017Cell}. This makes the total radiated power equal in all cases.

We consider a uniform linear array with omni-directional antennas. For links between the TUE and APs, we utilize the threeslope propagation model from \cite{bjornson2019making}. For links between the UAV and APs, the LoS and NLoS components are respectively modeled as \cite{bjornson2017massive,ozdogan2019massive}.
\begin{align}
\label{LoS} {{{\mathbf{\bar h}}}_l} &= \sqrt {\beta _l^{{\text{LoS}}}} {\left[ {1\;{e^{j2\pi {d_{\text{H}}}\sin \left( {{\varphi _l}} \right)}} \ldots {e^{j2\pi {d_{\text{H}}}\left( {N - 1} \right)\sin \left( {{\varphi _l}} \right)}}} \right]^{\text{T}}}, \\
\label{NLoS} {\left[ {{{\mathbf{R}}_l}} \right]_{s,m}} &= \frac{{\beta _l^{{\text{NLoS}}}}}{K}\sum\limits_{k = 1}^K {{e^{j\pi \left( {s - m} \right)\sin \left( {{\varphi _{l,k}}} \right) - \frac{{\sigma _\varphi ^2}}{2}{{\left( {\pi \left( {s - m} \right)\cos \left( {{\varphi _{l,k}}} \right)} \right)}^{^2}}}}} ,
\end{align}
where ${\beta _l^{{\text{LoS}}}}$ and ${\beta _l^{{\text{NLoS}}}}$ denote the large-scale fading coefficients. ${d_H} \leqslant 0.5$ is the antenna spacing parameter (in fractions of the wavelength). ${{\varphi _l}}$ is the angle of arrival (AoA) to the UAV seen from the AP, and ${\varphi _{l,k}} \sim \mathcal{U}\left[ {{\varphi _l} - {{40}^{\text{o}}},{\varphi _l} + {{40}^{\text{o}}}} \right]$ is the nominal AoA for the $k$ cluster. The multipath components of a cluster have Gaussian distributed AoAs, distributed around the nominal AoA with the angular standard deviation (ASD).
Moreover, the Rician factor is calculated as ${K_l} = 13 - 0.03{d_l}$ [dB], where $d_l = \sqrt{{{{\left\| {{{\mathbf{q}}_{\text{u}}}\left[ n \right] - {{\mathbf{q}}_l}} \right\|}^2} + {H^2}}}$ is the link distance.
Then, we can compute the large-scale fading parameters for the LoS and NLoS paths in \eqref{LoS} and \eqref{NLoS} respectively as
\begin{align}
\beta _l^{{\text{LoS}}} &= \frac{1}{N}{\left\| {{{{\mathbf{\bar h}}}_l}} \right\|^2} = \sqrt {{K_l}/\left( {{K_l} + 1} \right)} {\zeta _l}, \\
\beta _l^{{\text{NLoS}}} &= \frac{1}{N}{\text{tr}}\left( {{{\mathbf{R}}_l}} \right) = \sqrt {1/\left( {{K_l} + 1} \right)} {\zeta _l}.
\end{align}
It is worth noting that, in the following simulations, CF massive MIMO systems use a matched filtering receiver cooperation unless the LSFD receiver cooperation is specified.

\subsection{Uplink SE and downlink HE analysis}

In the following, we analyze the SE and HE of the considered systems from the time slot $1$ due to the initial pilot power at time slot $0$ comes from UAV.
The effect of spatial correlation is small because the short distance between AP and UAV leads to the strong LoS component in our considered systems.
We assume the spatial correlation $\text{ASD}=10^\text{o}$ and 5000 random realizations of the APs and UAV locations are generated.

\begin{figure}[t]
\begin{minipage}[t]{0.48\linewidth}	
\centering
\includegraphics[scale=0.55]{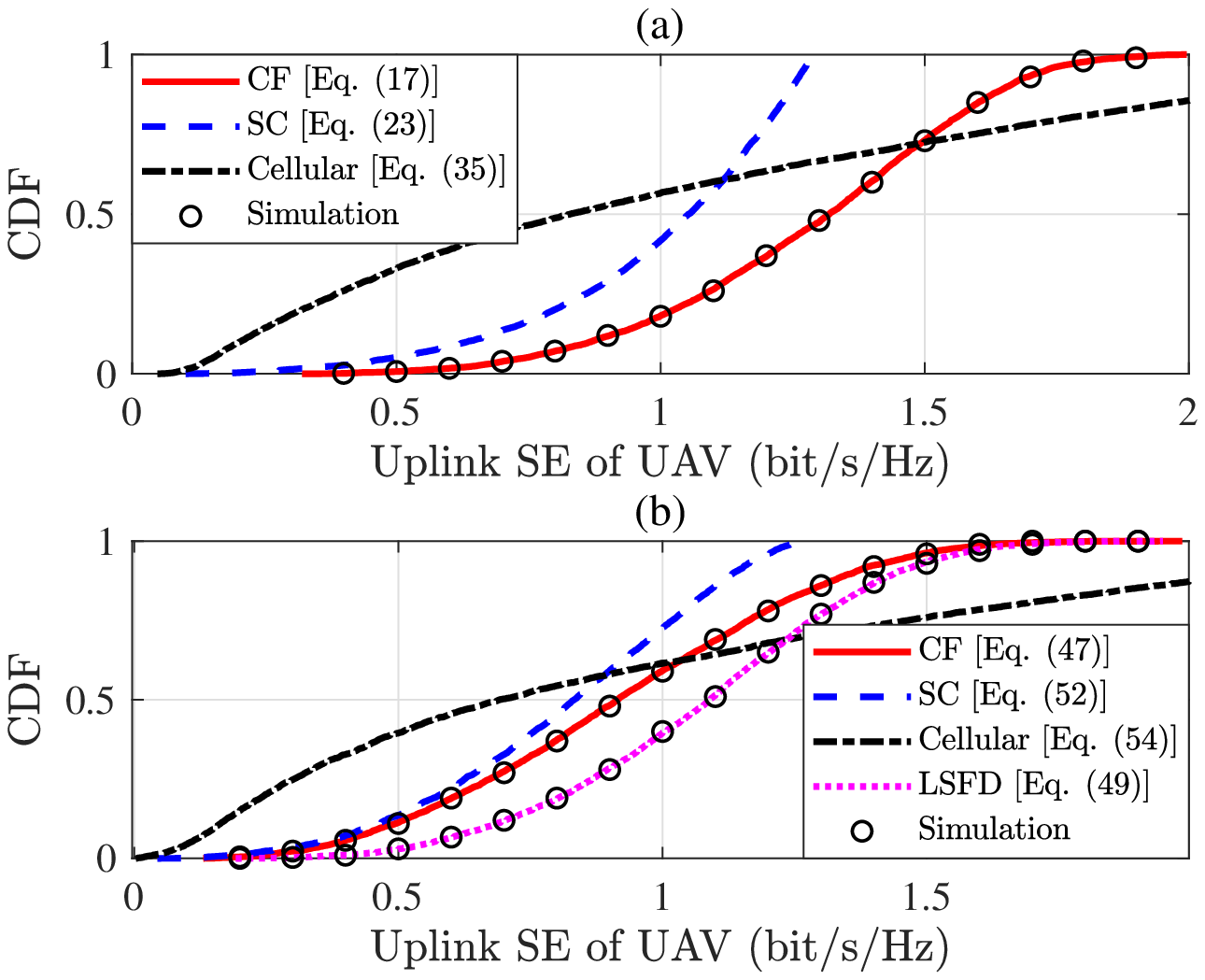}
\caption{CDF of uplink SE of UAV in CF massive MIMO, SC and cellular massive MIMO systems ($L\!=\!20$, $N\!=\!2$, $H\!=\!20$ m, $\kappa\!=\!0.98$, $\rho\!=\!0.5$). (a) Without TUE; (b) With TUE.} \vspace{-4mm}
\label{SE_three}
\end{minipage}
\hfill
\begin{minipage}[t]{0.48\linewidth}
\centering
\includegraphics[scale=0.55]{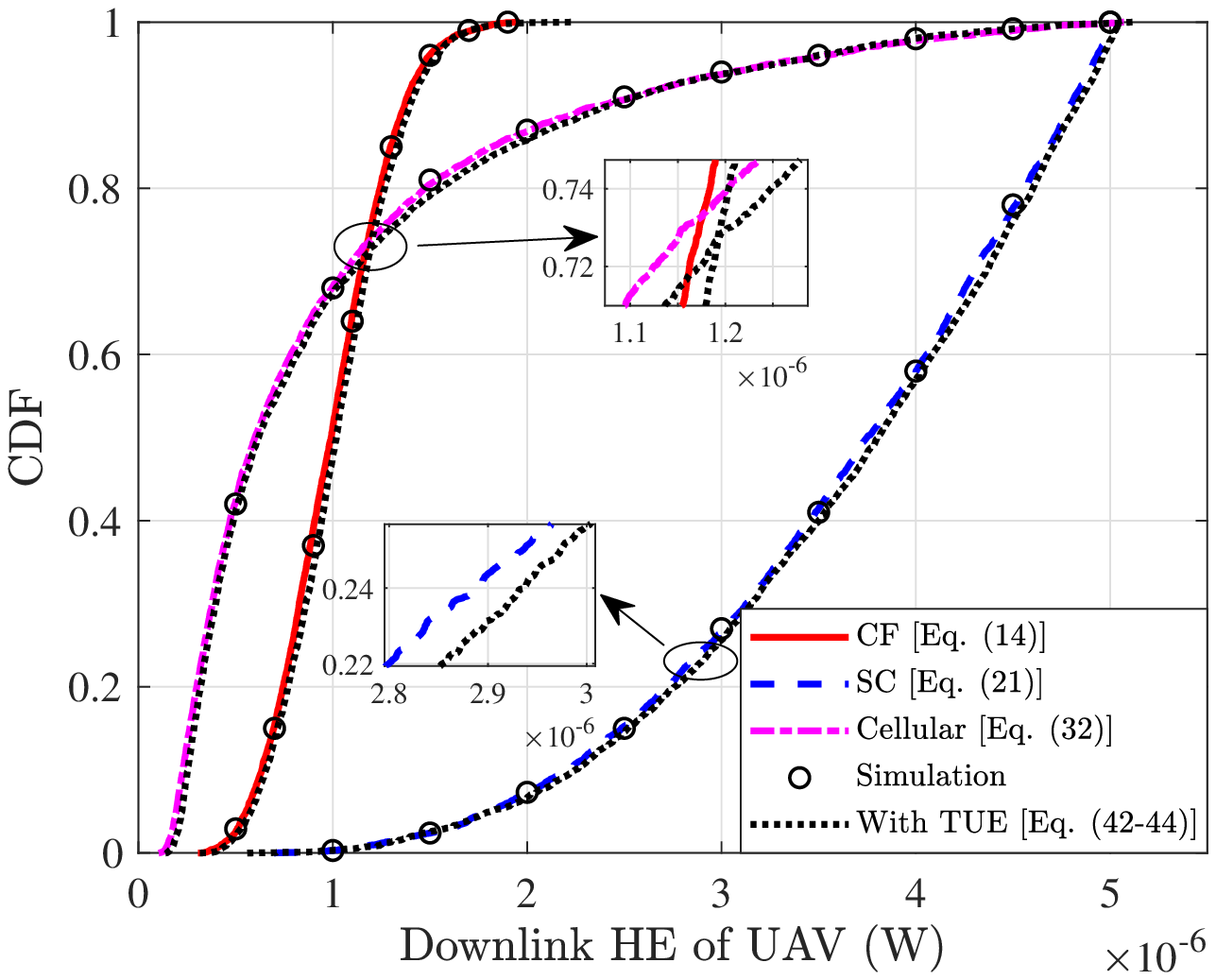}
\caption{CDF of downlink HE of UAV in CF massive MIMO, SC and cellular massive MIMO systems ($L=20$, $N=2$, $H=20$ m, $\kappa=0.98$, $\rho=0.5$).} \vspace{-4mm}
\label{HE_three}
\end{minipage}
\end{figure}

\begin{figure}[t]
\begin{minipage}[t]{0.48\linewidth}	
\centering
\includegraphics[scale=0.55]{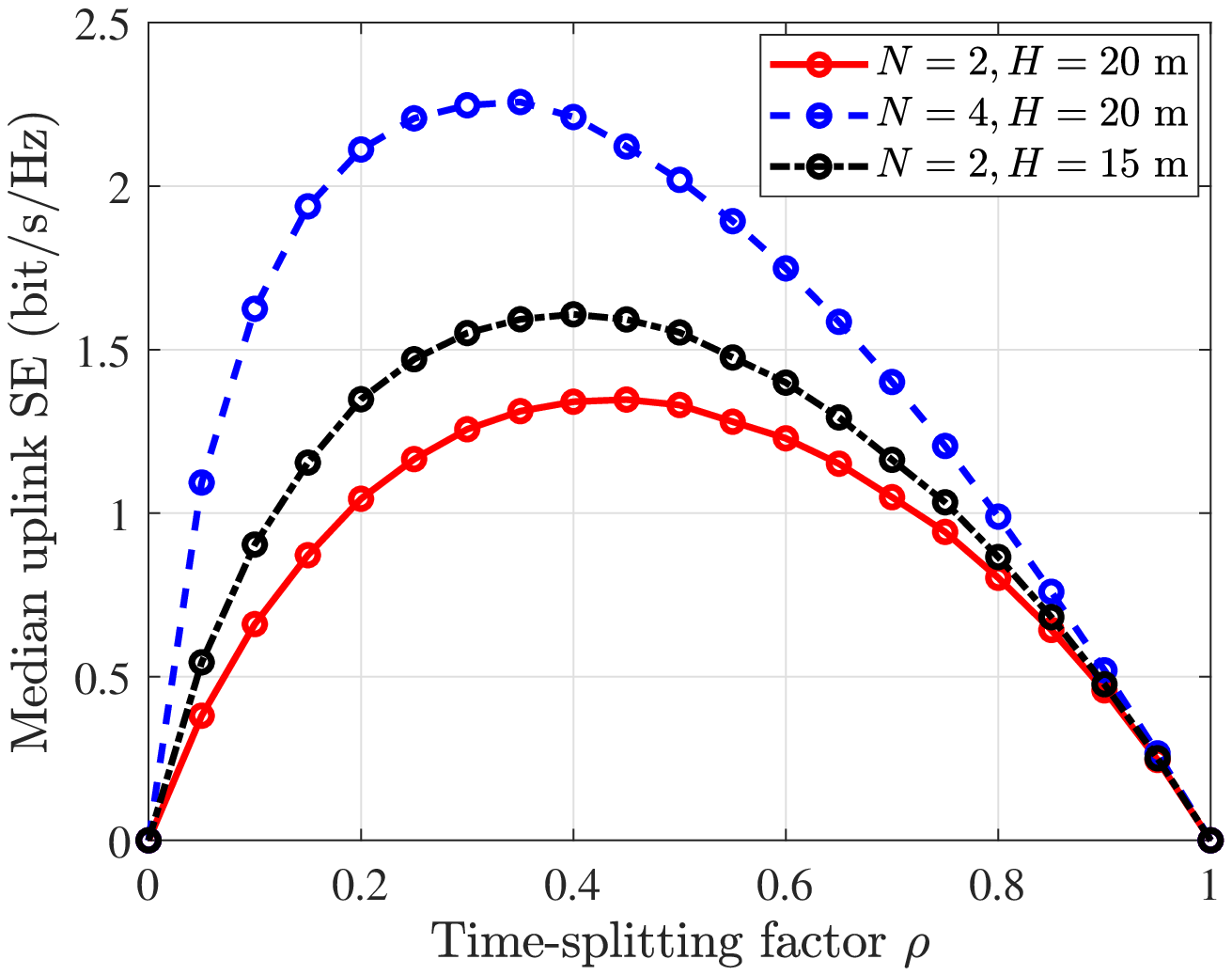}
\caption{Median uplink SE of UAV against the value of $\rho$ for CF massive MIMO with different values of $N$ and $H$ ($L=20$, $\kappa=0.98$).} \vspace{-4mm}
\label{rho_N_H}
\end{minipage}
\hfill
\begin{minipage}[t]{0.48\linewidth}
\centering
\includegraphics[scale=0.55]{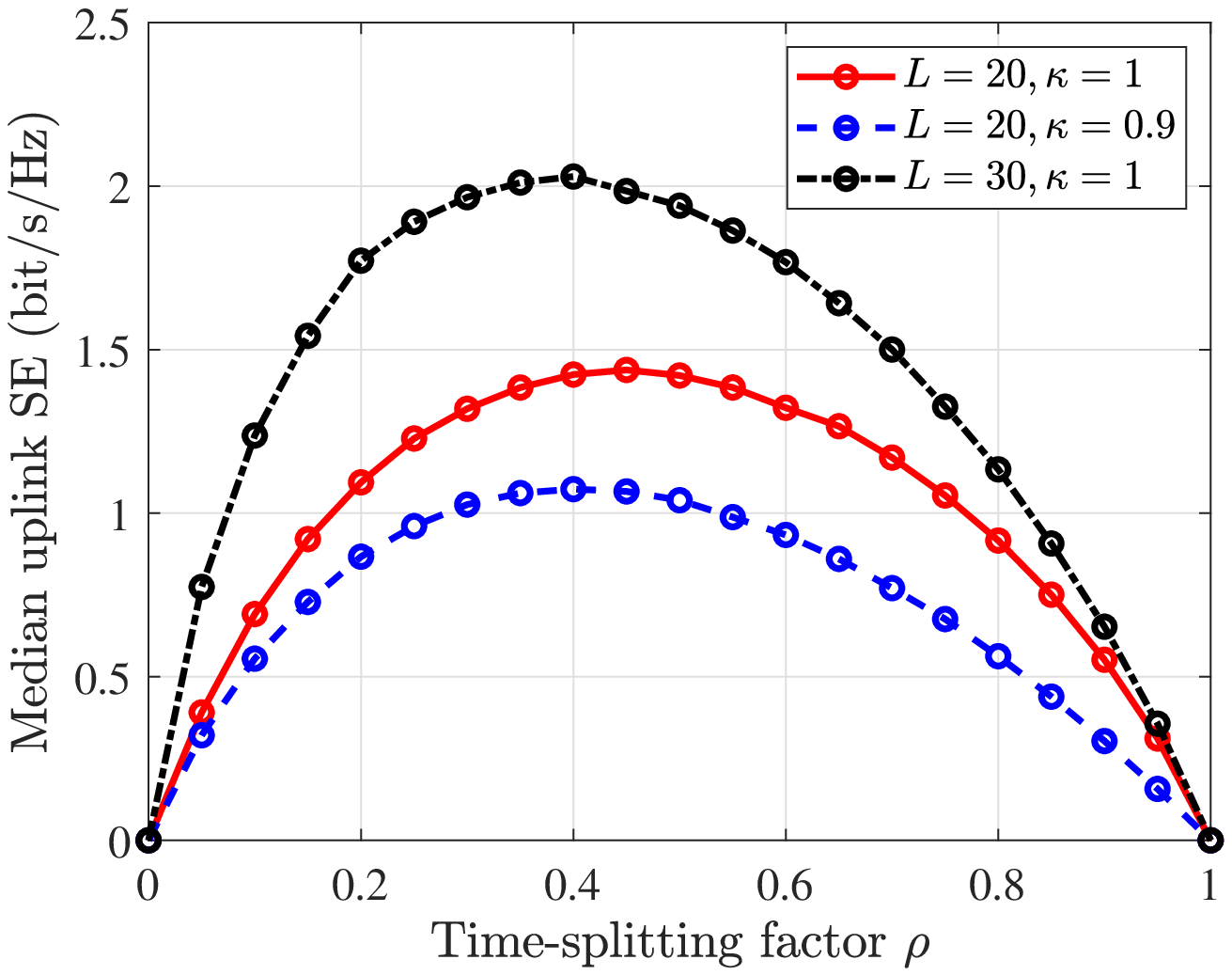}
\caption{Median uplink SE of UAV against the value of $\rho$ for CF massive MIMO with different values of $L$ and $\kappa$ ($N\!\!=\!\!2$, $H\!\!=\!\!20$ m).} \vspace{-4mm}
\label{rho_kappa_L}
\end{minipage}
\end{figure}

Fig.~\ref{SE_three} compares the CDF of the uplink SE of UAV achieved in the CF massive MIMO, SC and cellular massive MIMO systems for $\kappa=0.98$, $\rho=0.5$, respectively.
It is clear that the CF massive MIMO achieves two-fold and five-fold gain over cellular massive MIMO in terms of median and 95\%-likely uplink SE, respectively. The reason is that the average distance among APs and UAV hugely reduces in CF massive MIMO systems. It is beneficial for APs and UAV to obtain macro diversity gain. In addition, distributed antenna location and cooperative signal processing in CF massive MIMO provide more uniform coverage for UAV. Although a great SE gain can be achieved in cellular massive MIMO while UAV is close to the cellular BS, it has poor SE performance in far locations, especially in the edge of the cell. Furthermore, CF massive MIMO performs better than SC systems. The reason is that the UAV is only served by the SE-maximizing AP in SC systems. It means that only $N$ antennas are used to receive signal in uplink data transmission, which achieves a poor antenna diversity gain.
Moreover, we find that TUE has a bad effect on the SE performance of UAV, and the LSFD receiver cooperation can effectively reduce the interference.

The CDF of downlink HE of UAV in CF massive MIMO, SC and cellular massive MIMO systems is shown in Fig.~\ref{HE_three}, respectively. We also find that CF massive MIMO achieves two-fold and three-fold gain over cellular massive MIMO in terms of median and 95\%-likely downlink HE, respectively. The reason is that CF massive MIMO brings the antennas closer to UAV. It is helpful for UAV to harvest energy. For a fair comparison between CF massive MIMO and SC systems, $p_{\text{d}}^{{\text{sc}}}\left[ n \right] = Lp_{\text{d}}^{{\text{cf}}}\left[ n \right]$ is adopted \cite{Ngo2017Cell}. Therefore, we can find that SC systems obtain four-fold gain over CF massive MIMO in terms of median downlink HE. However, we notice that the uplink SE of SC is still smaller than CF massive MIMO. It is worth noting that, in practice, the transmit power of SC systems is not multiplied by the number of APs. Therefore, the performance of SC systems is poor in reality.
We further find that it is hard for the UAV to make use of the energy from TUE. The reason is that there is no beamforming operation and LoS link for TUE.

Fig.~\ref{rho_N_H} depicts the median uplink SE of UAV against the value of $\rho$ for CF massive MIMO with different values of $N$ and $H$. We set $L=20$ and $\kappa=0.98$. It is clear that the median uplink SE of UAV first increases and then decreases when changing the time-splitting factor $\rho$ from $0$ to $1$. Because $\rho=0$ and $\rho=1$ mean there are no energy harvesting and no information transmission, respectively. They both can result in the communication failure completely. In addition, increasing the number of antennas per AP and decreasing the flight altitude of UAV both can improve the SE of the considered systems. Furthermore, smaller $\rho$ is beneficial for the optimal operating point of SE in both systems. The reason is that more antennas and shorter distance make a larger gain on downlink energy harvesting than uplink data transmission. Therefore, smaller $\rho$ is needed to achieve balance.

The median uplink SE of UAV against the value of $\rho$ for CF massive MIMO with different values of $L$ and $\kappa$ is illustrated in Fig.~\ref{rho_kappa_L}. We set the number of antennas per AP $N=2$ and the flight altitude of UAV $H=20$ m. As we expected, increasing the number of APs leads to a higher median SE. Smaller time of energy harvesting is beneficial for the optimal operating point of SE. For the case of $\rho=0.4$ and the hardware quality factor of UAV $\kappa$ reduces from $1$ to $0.9$, the median SE has a $25\%$ loss. Moreover, we notice that hardware impairments of UAV require longer time of data transmission are preferred for the optimal operating point of SE. The reason is that the hardware impairments at the UAV has a greater performance loss than the ones at the APs. This finding is consistent with \cite{zheng2020efficient}. Therefore, increasing the hardware impairments at UAV makes a larger loss on uplink data transmission than downlink energy harvesting. Larger length of data transmission is needed to achieve balance.

\subsection{UAV trajectory analysis}

In the following, we analyze the trajectory design schemes of UAV and per-slot SE. We consider spatially correlated Rician fading with $\text{ASD}=10^\text{o}$ and 200 random realizations of the APs and TUE locations are generated for per-slot SE analysis.

\begin{figure}[t]
\begin{minipage}[t]{0.48\linewidth}	
\centering
\includegraphics[scale=0.55]{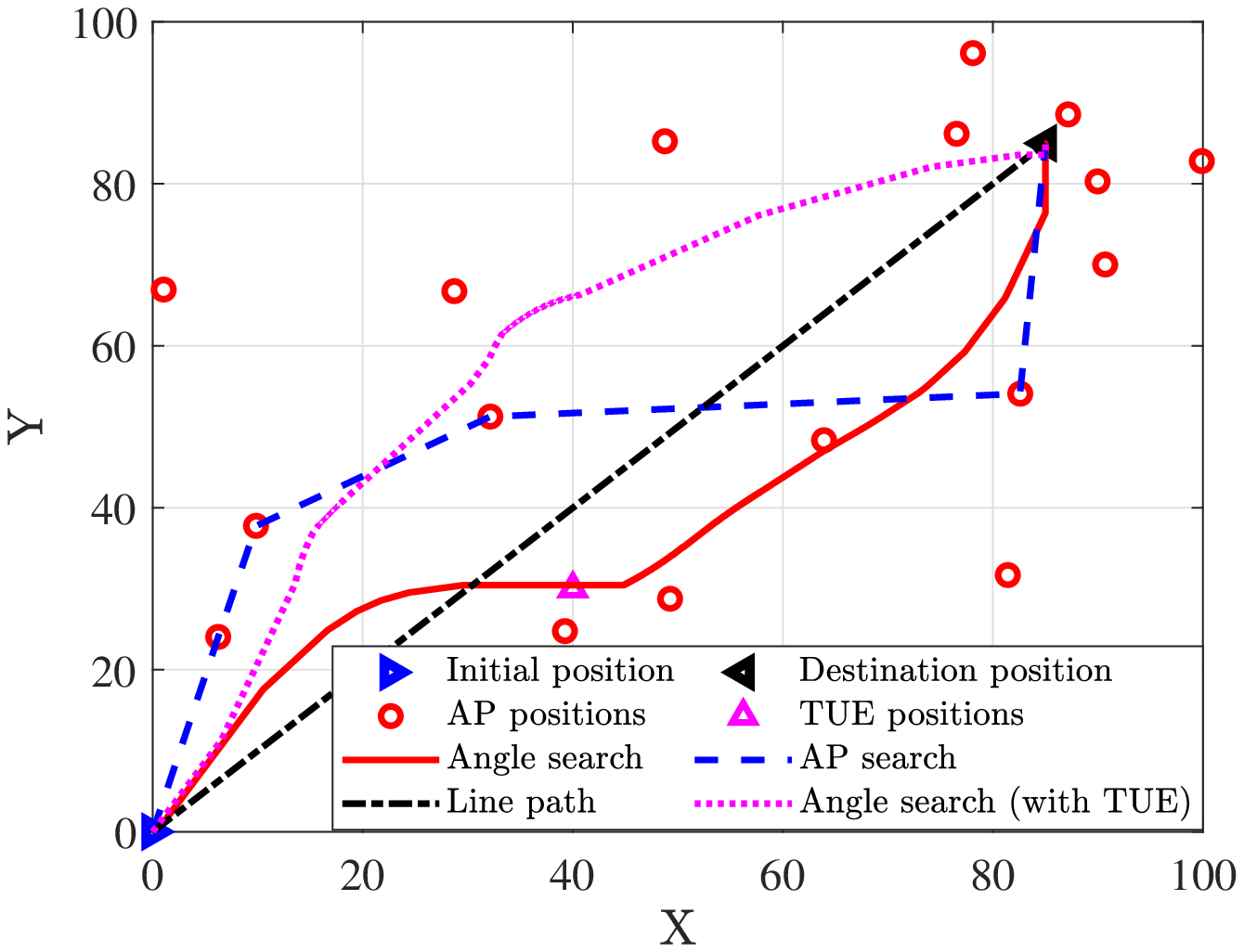}
\caption{UAV Trajectory for angle search, AP search and line path flight schemes in CF massive MIMO systems ($L=20$, $N=2$, $H=20$ m, $\kappa=0.98$, $\rho=0.5$).} \vspace{-4mm}
\label{trajecory1}
\end{minipage}
\hfill
\begin{minipage}[t]{0.48\linewidth}
\centering
\includegraphics[scale=0.55]{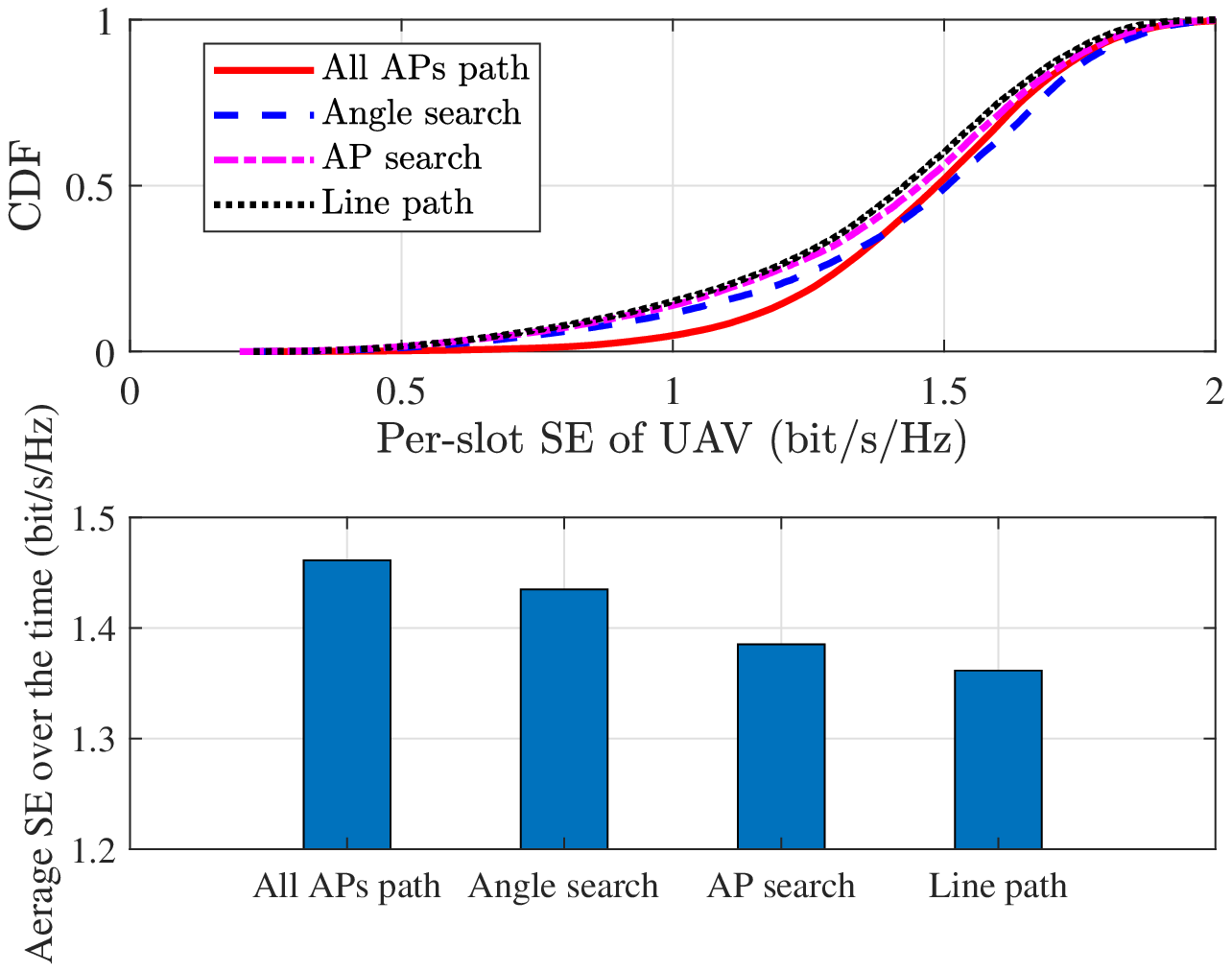}
\caption{SE performance of UAV for angle search, AP search, line path flight and all APs path schemes in CF massive MIMO systems ($L\!=\!20$, $N\!=\!2$, $H\!=\!20$ m, $\kappa\!=\!0.98$, $\rho\!=\!0.5$).} \vspace{-4mm}
\label{tabu_nearAP_line}
\end{minipage}
\end{figure}

\begin{figure}[t]
\begin{minipage}[t]{0.48\linewidth}	
\centering
\includegraphics[scale=0.55]{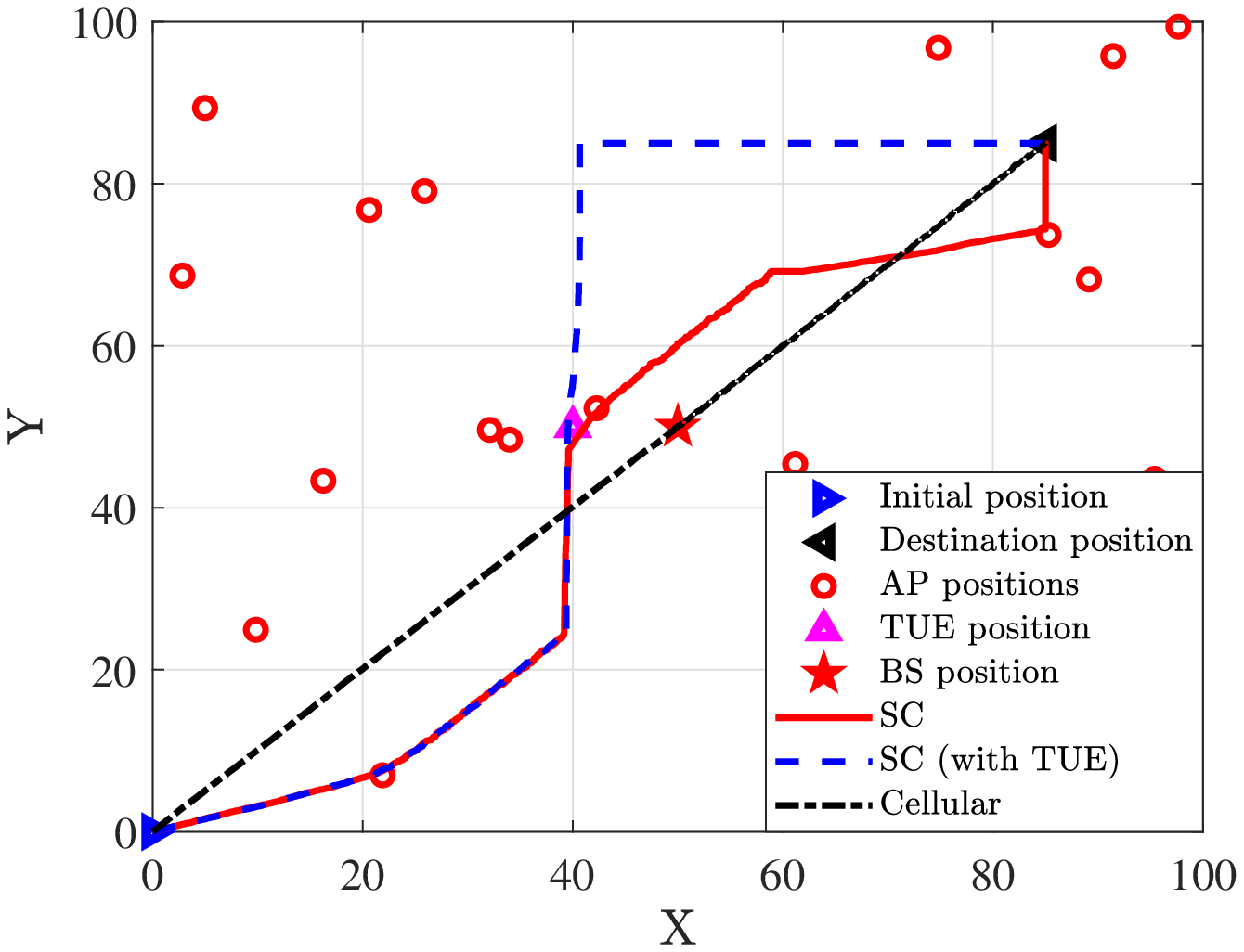}
\caption{UAV Trajectory for SC and cellular massive MIMO systems with the angle search flight scheme ($L=20$, $N=2$, $H=20$ m, $\kappa=0.98$, $\rho=0.5$).} \vspace{-4mm}
\label{trajecory2}
\end{minipage}
\hfill
\begin{minipage}[t]{0.48\linewidth}
\centering
\includegraphics[scale=0.55]{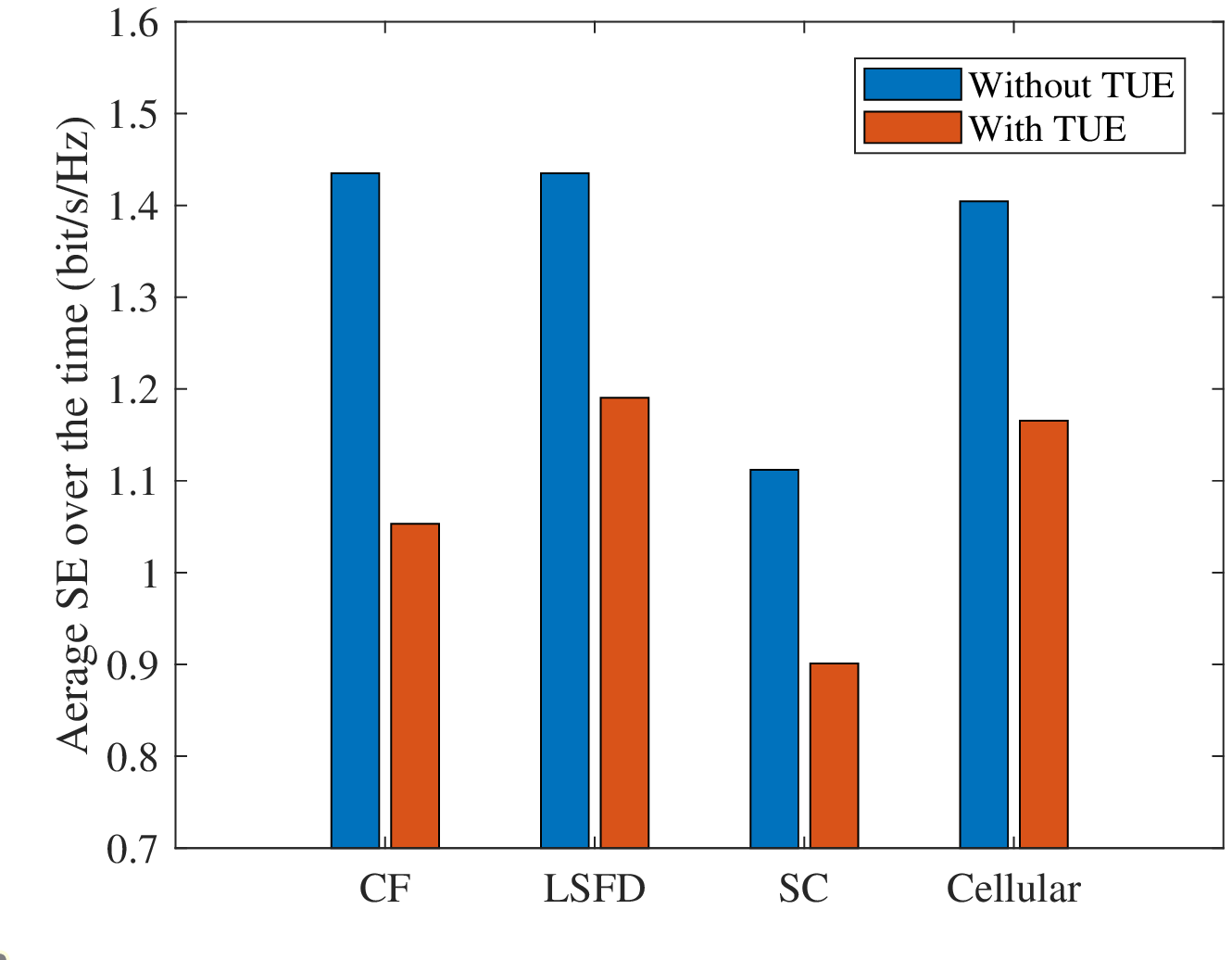}
\caption{Average SE of UAV for CF, LSFD, SC and cellular massive MIMO systems with the angle search flight scheme ($L=20$, $N=2$, $H=20$ m, $\kappa=0.98$, $\rho=0.5$).} \vspace{-4mm}
\label{CF_SC_Cellular}
\end{minipage}
\end{figure}

Fig.~\ref{trajecory1} shows the trajectory of UAV for angle search, AP search and line path trajectory design schemes in CF massive MIMO systems, respectively. We assume $L=20$ APs are randomly generated within a $100\text{m}\times100\text{m}$ square, the initial UAV position is located in ${{\mathbf{q}}_{\text{u}}}\left[ 0 \right] = \left[ {0,0} \right]$, the UAV destination is located in ${{\mathbf{q}}_{{\text{des}}}} = \left[ {85,85} \right]$, and the number of neighbouring positions $M=10$ in angle search scheme\footnote{Because the minimum flight distance $d_{\text{min}}=4$ cm is small, angle search interval $10^\text{o}$ ($M=10$) is very accurate.}. It is clear from Fig.~\ref{trajecory1} that the angle search trajectory design scheme can get more dense APs around the UAV flight path. The trajectory obtained by AP search scheme flies over four APs, but it is far away from other APs. UAV with line path scheme only be close to one AP.
In addition, when there is a TUE, the angle search scheme can bypass the APs with large interference. However, AP search and line path schemes remain unchanged.
Due to these three schemes have different flight time, we apply per-slot SE and average SE over time to characterize performance differences in the following.

The CDF of uplink per-slot SE of UAV for angle search, AP search, line path and all APs path flight schemes in CF massive MIMO are compared in Fig.~\ref{tabu_nearAP_line}, respectively. It is interesting to find that, compared with the line path flight scheme, angle search and AP search trajectory design schemes have $5\%$ and $2\%$ gain in terms of average SE over time, respectively. It is worth noting that CF massive MIMO provides ubiquitous and uniform coverage, and it is a challenge to substantially improve the performance of CF massive MIMO by designing the trajectory of UAV.
In addition, all APs path scheme is not constrained by the target direction, so it can achieve a higher average SE performance. However, the flight time of all APs path scheme is far greater than other three schemes, which is impractical considering the limited flight power of UAVs.

Fig.~\ref{trajecory2} illustrates the trajectory of UAV for SC and cellular massive MIMO systems with the angle search flight scheme, respectively. We also use $L=20$ APs randomly distributed within a $100\text{m}\times100\text{m}$ square. The initial UAV position is ${{\mathbf{q}}_{\text{u}}}\left[ 0 \right] = \left[ {0,0} \right]$ and the UAV destination is ${{\mathbf{q}}_{{\text{des}}}} = \left[ {85,85} \right]$.
In addition, for cellular massive MIMO, the BS with $LN$ antennas is located in ${{\mathbf{q}}_{{\text{bs}}}} = \left[ {50,50} \right]$. We find that the UAV flies over the serving AP and BS both in SC and cellular massive MIMO systems, because it is only served by one transmitter.
We also find that, when there is a great interference in one cell, the UAV will avoid being served by the AP in this cell. However, UAV trajectory remains unchanged in the cellular system, because the UAV is only served by one BS.

The average SE performance of UAV for CF, SC and cellular massive MIMO systems with the angle search flight scheme is shown in Fig.~\ref{CF_SC_Cellular}, respectively.
It is clear that, with the angle search flight scheme, the average SE performance of UAV in CF massive MIMO system performs better than it in both SC and cellular massive MIMO systems. However, when there is a TUE interference, UAV in cellular massive MIMO system has a larger average SE performance than both it in CF and SC systems. According to the analysis of Fig.~\ref{SE_three}, only close to the BS, UAV in cellular systems can achieve great SE. Therefore, cellular massive MIMI systems can not provide uniform SE performance over the flight time of UAV as the CF massive MIMO system does. In addition, LSFD receiver cooperation can be applied in CF systems to get higher average SE performance than cellular systems. Note that, only while the TUE interference exists, LSFD receiver cooperation does work.

\section{Conclusions}\label{se:conclusion}
In this paper, we investigate the downlink HE and uplink SE of CF massive MIMO taking the effects of both WPT and hardware impairments in account for UAV communications. The UAV harvests energy from uplink APs' energy symbols, then it is used to support the uplink data and pilot transmission. In addition, we analyse SC and cellular massive MIMO systems for comparison. Based on the channel estimates, we derive novel closed-form downlink HE and uplink SE expressions and quantify the impact of hardware impairments for considered systems. It is important to see that, compared with SC and cellular massive MIMO, CF massive MIMO performs best, and the LSFD receiver cooperation can reduce the interference of TUE. Moreover, the maximum SE can be obtained by changing the time-splitting fraction $\rho$, the number of antennas, the altitude of UAV, and the hardware quality of UAVs. Furthermore, we propose an angle search trajectory design scheme to improve the per-slot SE. The AP search and line path trajectory design schemes are also analyzed for comparison. A large performance gain in terms of SE can be achieved by our angle search scheme. Moreover, the angle search scheme can bypass the APs with large interference in both CF massive MIMO and SC systems.
In future work, we will investigate the spatial correlated multiple-antennas UAV communication in CF massive MIMO systems. Moreover, the power control and interference coordination of multiple UAVs also will be considered to improve the performance of systems.

\begin{appendices}
\section{Proof of Theorem 1}

Submitting the beamforming scalar ${\mathbf{w}}_l\left[ n \right] = {{{{{\mathbf{\hat g}}}_l}}}\left[ n \right]/{{\sqrt {\mathbb{E}\left\{ {{{\left\| {{{{\mathbf{\hat g}}}_l}}\left[ n \right] \right\|}^2}} \right\}} }}$ in \eqref{P_in}, we can write the received energy $P_\text{in}\left[ n \right]$ as
\begin{align}
{P_{\text{in}}}\left[ n \right] = {p_{\text{d}}}\left[ n \right]\sum\limits_{l = 1}^L {\frac{{\mathbb{E}\left\{ {{{\left| {{\mathbf{g}}_l^{\text{H}}\left[ n \right]{{{\mathbf{\hat g}}}_l}}\left[ n \right] \right|}^2}} \right\}}}{{\mathbb{E}\left\{ {{{\left\| {{{{\mathbf{\hat g}}}_l}}\left[ n \right] \right\|}^2}} \right\}}}} .
\end{align}
Based on the properties of LMMSE estimation \cite{bjornson2017massive}, ${{{\mathbf{\hat g}}}_l}\left[ n \right]$ and ${{{\mathbf{\tilde g}}}_l}\left[ n \right]$ are independent. We have
\begin{align}\label{g_ghat}
  \mathbb{E}\left\{ {{\mathbf{g}}_l^{\text{H}}\left[ n \right]{{{\mathbf{\hat g}}}_l}}\left[ n \right] \right\} &= \mathbb{E}\left\{ {{{\left( {{{{\mathbf{\hat g}}}_l}\left[ n \right] + {{{\mathbf{\tilde g}}}_l}}\left[ n \right] \right)}^{\text{H}}}{{{\mathbf{\hat g}}}_l}}\left[ n \right] \right\} = \mathbb{E}\left\{ {{{\left\| {{{{\mathbf{\hat g}}}_l}}\left[ n \right] \right\|}^2}} \right\} \notag \\
   &= \kappa p\left[ n \right]{\text{tr}}\left( {{{\mathbf{R}}_l}\left[ n \right]{{\mathbf{\Psi }}_l}\left[ n \right]{{\mathbf{R}}_l}}\left[ n \right] \right) + {\left\| {{{{\mathbf{\bar h}}}_l}}\left[ n \right] \right\|^2} = {\text{tr}}\left( {{{\mathbf{Q}}_l}}\left[ n \right] \right) + {\left\| {{{{\mathbf{\bar h}}}_l}}\left[ n \right] \right\|^2}.
\end{align}
In addition, we have
\begin{align}\label{gghat}
  \mathbb{E}\left\{ {{{\left| {{\mathbf{g}}_l^{\text{H}}\left[ n \right]{{{\mathbf{\hat g}}}_l}}\left[ n \right] \right|}^2}} \right\} &= \mathbb{E}\left\{ {{{\left| {{{\left( {{{{\mathbf{\hat g}}}_l}\left[ n \right] + {{{\mathbf{\tilde g}}}_l}}\left[ n \right] \right)}^{\text{H}}}{{{\mathbf{\hat g}}}_l}}\left[ n \right] \right|}^2}} \right\} \notag\\
  &= \mathbb{E}\left\{ {{{\left| {{\mathbf{\hat g}}_l^{\text{H}}\left[ n \right]{{{\mathbf{\hat g}}}_l}}\left[ n \right] \right|}^2}} \right\} + \mathbb{E}\left\{ {{{\left| {{\mathbf{\tilde g}}_l^{\text{H}}\left[ n \right]{{{\mathbf{\hat g}}}_l}}\left[ n \right] \right|}^2}} \right\}.
\end{align}
In order to derive the closed-form expression for \eqref{gghat}, note that ${{{\mathbf{\hat g}}}_l}\left[ n \right] = {\mathbf{Q}}_l^{\frac{1}{2}}\left[ n \right]{\mathbf{m}} + {{{\mathbf{\bar h}}}_l}\left[ n \right]$ where ${\mathbf{m}} \sim \mathcal{C}\mathcal{N}\left( {{\mathbf{0}},{{\mathbf{I}}_N}} \right)$. Therefore,
\begin{align}\label{c0}
\!\!\!\!\mathbb{E}\!\left\{ {{{\left| {{\mathbf{\hat g}}_l^{\text{H}}\left[ n \right]{{{\mathbf{\hat g}}}_l}}\left[ n \right] \right|}^2}} \right\} &\!=\!
  \mathbb{E}\left\{ {\left| {\underbrace {{{\mathbf{m}}^{\text{H}}}{{\left( {{\mathbf{Q}}_l^{\text{H}}\left[ n \right]} \right)}^{\frac{1}{2}}}{\mathbf{Q}}_l^{\frac{1}{2}}\left[ n \right]{\mathbf{m}}}_{a\left[ n \right]}} \right.} \right. \notag \\
  &\left. {{{\left. { + \underbrace {{{\mathbf{m}}^{\text{H}}}{{\left( {{\mathbf{Q}}_l^{\text{H}}\left[ n \right]} \right)}^{\frac{1}{2}}}{{{\mathbf{\bar h}}}_l}\left[ n \right]}_{b\left[ n \right]} + \underbrace {{\mathbf{\bar h}}_l^{\text{H}}\left[ n \right]{\mathbf{Q}}_l^{\frac{1}{2}}\left[ n \right]{\mathbf{m}}}_{c\left[ n \right]} + \underbrace {{\mathbf{\bar h}}_l^{\text{H}}\left[ n \right]{{{\mathbf{\bar h}}}_l}\left[ n \right]}_{d\left[ n \right]}} \right|}^2}} \right\} .
\end{align}
We can compute each term as
\begin{align}
\label{c1} \mathbb{E}\left\{ {a\left[ n \right]{a^*}\left[ n \right]} \right\} &= \mathbb{E}\left\{ {{{\left| {{{\mathbf{m}}^{\text{H}}}{{\left( {{\mathbf{Q}}_l^{\text{H}}}\left[ n \right] \right)}^{\frac{1}{2}}}{\mathbf{Q}}_l^{\frac{1}{2}}\left[ n \right]{\mathbf{m}}} \right|}^2}} \right\} \notag\\
 &= {\left| {{\text{tr}}\left( {{{\left( {{\mathbf{Q}}_l^{\text{H}}}\left[ n \right] \right)}^{\frac{1}{2}}}{\mathbf{Q}}_l^{\frac{1}{2}}}\left[ n \right] \right)} \right|^2} + {\text{tr}}\left( {{{\mathbf{Q}}_l}\left[ n \right]{{\mathbf{Q}}_l}}\left[ n \right] \right), \\
\label{c2} \mathbb{E}\left\{ {b\left[ n \right]{b^*}\left[ n \right]} \right\} &= \mathbb{E}\left\{ {c\left[ n \right]{c^*}\left[ n \right]} \right\} \notag\\
&= \mathbb{E}\left\{ {{\mathbf{\bar h}}_l^{\text{H}}\left[ n \right]{\mathbf{Q}}_l^{\frac{1}{2}}\left[ n \right]{\mathbf{m}}{{\mathbf{m}}^{\text{H}}}{{\left( {{\mathbf{Q}}_l^{\text{H}}}\left[ n \right] \right)}^{\frac{1}{2}}}{{{\mathbf{\bar h}}}_l}}\left[ n \right] \right\} = {\mathbf{\bar h}}_l^{\text{H}}\left[ n \right]{{\mathbf{Q}}_l}\left[ n \right]{{{\mathbf{\bar h}}}_l}\left[ n \right], \\
\label{c3} \mathbb{E}\left\{ {d\left[ n \right]{d^*}\left[ n \right]} \right\} &= {\left| {{\mathbf{\bar h}}_l^{\text{H}}\left[ n \right]{{{\mathbf{\bar h}}}_l}}\left[ n \right] \right|^2}, \\
\label{c4} \mathbb{E}\left\{ {a\left[ n \right]{d^*}\left[ n \right]} \right\} &= \mathbb{E}\left\{ {d\left[ n \right]{a^*}\left[ n \right]} \right\} = {\text{tr}}\left( {{{\left( {{\mathbf{Q}}_l^{\text{H}}}\left[ n \right] \right)}^{\frac{1}{2}}}{\mathbf{Q}}_l^{\frac{1}{2}}}\left[ n \right] \right){\mathbf{\bar h}}_l^{\text{H}}\left[ n \right]{{{\mathbf{\bar h}}}_l}\left[ n \right].
\end{align}
Other terms are zero due to the circular symmetry property of ${\mathbf{m}}$. With the help of \eqref{c0}, \eqref{c1}, \eqref{c2}, \eqref{c3}, \eqref{c4} and \eqref{Q}, we have
\begin{align}\label{a1}
  &\mathbb{E}\left\{ {{{\left| {{\mathbf{\hat g}}_l^{\text{H}}\left[ n \right]{{{\mathbf{\hat g}}}_l}}\left[ n \right] \right|}^2}} \right\} = {\kappa ^2}{p^2}\left[ n \right]{\left| {{\text{tr}}\left( {{{\mathbf{R}}_l}\left[ n \right]{{\mathbf{\Psi }}_l}\left[ n \right]{{\mathbf{R}}_l}}\left[ n \right] \right)} \right|^2} + {\left| {{\mathbf{\bar h}}_l^{\text{H}}\left[ n \right]{{{\mathbf{\bar h}}}_l}}\left[ n \right] \right|^2} \notag\\
   &+ 2\kappa p\left[ n \right]\operatorname{Re} \left\{ {{\text{tr}}\left( {{{\mathbf{R}}_l}\left[ n \right]{{\mathbf{\Psi }}_l}\left[ n \right]{{\mathbf{R}}_l}}\left[ n \right] \right){\mathbf{\bar h}}_l^{\text{H}}\left[ n \right]{{{\mathbf{\bar h}}}_l}}\left[ n \right] \right\} + \kappa p\left[ n \right]{\mathbf{\bar h}}_l^{\text{H}}\left[ n \right]{{\mathbf{R}}_l}\left[ n \right]{{\mathbf{\Psi }}_l}\left[ n \right]{{\mathbf{R}}_l}\left[ n \right]{{{\mathbf{\bar h}}}_l}\left[ n \right] \notag\\
   &+ \kappa p\left[ n \right]{\text{tr}}\left( {\left( {{{\mathbf{R}}_l}\left[ n \right] - {{\mathbf{C}}_l}}\left[ n \right] \right){{\mathbf{R}}_l}\left[ n \right]{{\mathbf{\Psi }}_l}\left[ n \right]{{\mathbf{R}}_l}}\left[ n \right] \right) + {\mathbf{\bar h}}_l^{\text{H}}\left[ n \right]\left( {{{\mathbf{R}}_l}\left[ n \right] - {{\mathbf{C}}_l}}\left[ n \right] \right){{{\mathbf{\bar h}}}_l}\left[ n \right] .
\end{align}
Furthermore, we obtain
\begin{align}\label{a2}
  \mathbb{E}\left\{ {{{\left| {{\mathbf{\tilde g}}_l^{\text{H}}\left[ n \right]{{{\mathbf{\hat g}}}_l}}\left[ n \right] \right|}^2}} \right\} &= \kappa p\left[ n \right]{\text{tr}}\left( {{{\mathbf{C}}_l}\left[ n \right]{{\mathbf{R}}_l}\left[ n \right]{{\mathbf{\Psi }}_l}\left[ n \right]{{\mathbf{R}}_l}}\left[ n \right] \right) + {\mathbf{\bar h}}_l^{\text{H}}\left[ n \right]{{\mathbf{C}}_l}\left[ n \right]{{{\mathbf{\bar h}}}_l}\left[ n \right].
\end{align}
Using \eqref{a1} and \eqref{a2}, we can write \eqref{gghat} as
\begin{align}\label{g_ghat2}
&{\mathbb{E}\left\{ {{{\left| {{\mathbf{g}}_l^{\text{H}}\left[ n \right]{{{\mathbf{\hat g}}}_l}}\left[ n \right] \right|}^2}} \right\}} = {\kappa ^2}{p^2}\left[ n \right]{\left| {{\text{tr}}\left( {{{\mathbf{R}}_l}\left[ n \right]{{\mathbf{\Psi }}_l}\left[ n \right]{{\mathbf{R}}_l}}\left[ n \right] \right)} \right|^2}  + {\left| {{\mathbf{\bar h}}_l^{\text{H}}\left[ n \right]{{{\mathbf{\bar h}}}_l}}\left[ n \right] \right|^2} \notag \\
 &+ 2\kappa p\left[ n \right]\operatorname{Re} \left\{ {{\text{tr}}\left( {{{\mathbf{R}}_l}\left[ n \right]{{\mathbf{\Psi }}_l}\left[ n \right]{{\mathbf{R}}_l}}\left[ n \right] \right){\mathbf{\bar h}}_l^{\text{H}}\left[ n \right]{{{\mathbf{\bar h}}}_l}}\left[ n \right] \right\} + \kappa p\left[ n \right]{\text{tr}}\left( {{{\mathbf{R}}_l}\left[ n \right]{{\mathbf{R}}_l}\left[ n \right]{{\mathbf{\Psi }}_l}\left[ n \right]{{\mathbf{R}}_l}}\left[ n \right] \right) \notag\\
 &+ \kappa p\left[ n \right]{\mathbf{\bar h}}_l^{\text{H}}\left[ n \right]{{\mathbf{R}}_l}\left[ n \right]{{\mathbf{\Psi }}_l}\left[ n \right]{{\mathbf{R}}_l}\left[ n \right]{{{\mathbf{\bar h}}}_l}\left[ n \right] + {\mathbf{\bar h}}_l^{\text{H}}\left[ n \right]{{\mathbf{R}}_l}\left[ n \right]{{{\mathbf{\bar h}}}_l}\left[ n \right]
\end{align}
to finish the proof.

\section{Proof of Theorem 2}

Using the use-and-then-forget capacity bound in \cite{bjornson2017massive}, the SE of UAV at the time slot $n$ is
\begin{align}
{\text{SE}}\left[ n \right] =\frac{1-\tau_p-\tau_e}{\tau_c} \log \left( {1 + \frac{{\mathbb{E}\left\{ {{{\left| {{\text{DS}}}\left[ n \right] \right|}^2}} \right\}}}{{\mathbb{E}\left\{ {{{\left| {{\text{BU}}}\left[ n \right] \right|}^2}} \right\} + \mathbb{E}\left\{ {{{\left| {{\text{HI}}}\left[ n \right] \right|}^2}} \right\} + \mathbb{E}\left\{ {{{\left| {{\text{NS}}}\left[ n \right] \right|}^2}} \right\}}}} \right).
\end{align}
We compute each term of ${\text{SE}}\left[ n \right]$ to finish the proof.

\emph{1) Compute ${\mathbb{E}\left\{ {{{\left| {{\mathrm{DS}}}\left[ n \right] \right|}^2}} \right\}}$:}
With the help of \eqref{g_ghat}, we obtain
\begin{align}
\mathbb{E}\left\{ {{{\left| {{\text{DS}}}\left[ n \right] \right|}^2}} \right\} &= \mathbb{E}\left\{ {{{\left| {\sqrt {\kappa {p_{\text{u}}}\left[ n \right]} \sum\limits_{l = 1}^L {\mathbb{E}\left\{ {{\mathbf{\hat g}}_l^{\text{H}}\left[ n \right]{{\mathbf{g}}_l}}\left[ n \right] \right\}s\left[ n \right]} } \right|}^2}} \right\} \notag\\
&= \kappa {p_{\text{u}}}\left[ n \right]{\left| {\sum\limits_{l = 1}^L {\left( {\kappa p\left[ n \right]{\text{tr}}\left( {{{\mathbf{R}}_l}\left[ n \right]{{\mathbf{\Psi }}_l}\left[ n \right]{{\mathbf{R}}_l}}\left[ n \right] \right) + {{\left\| {{{{\mathbf{\bar h}}}_l}}\left[ n \right] \right\|}^2}} \right)} } \right|^2} \notag\\
&= \kappa {p_{\text{u}}}\left[ n \right]{\left| {\sum\limits_{l = 1}^L {\left( {{\text{tr}}\left( {{{\mathbf{Q}}_l}}\left[ n \right] \right) + {{\left\| {{{{\mathbf{\bar h}}}_l}}\left[ n \right] \right\|}^2}} \right)} } \right|^2} .
\end{align}

\emph{2) Compute ${\mathbb{E}\left\{ {{{\left| {{\mathrm{BU}}}\left[ n \right] \right|}^2}} \right\}}$:}
Using \cite[Eq. (28)]{zheng2020efficient}, we have
\begin{align}\label{BU1}
\mathbb{E}\!\left\{ {{{\left| {{\text{BU}}}\!\left[ n \right] \right|}^2}} \right\} \! &=\kappa {p_{\text{u}}\left[ n \right]}\!\left(\! \sum\limits_{l = 1}^L {\mathbb{E}\!\left\{ {{{\left| {{\mathbf{\hat g}}_l^{\text{H}}\left[ n \right]{{\mathbf{g}}_l}\left[ n \right]} \right|}^2}} \right\}}  \!+\! \sum\limits_{l = 1}^L {\sum\limits_{m \ne l}^L {\mathbb{E}\!\left\{ {\mathbf{\hat g}}_l^{\text{H}}\left[ n \right]{{\mathbf{g}}_l}\left[ n \right]\right\}\!\mathbb{E}\!\left\{{\mathbf{\hat g}}_m^{\text{H}}\left[ n \right]{{\mathbf{g}}_m}\left[ n \right] \right\}} } \!\right) \notag\\
&-\kappa {p_{\text{u}}}\left[ n \right]{\left| {\sum\limits_{l = 1}^L {\mathbb{E}\left\{ {{\mathbf{\hat g}}_l^{\text{H}}\left[ n \right]{{\mathbf{g}}_l}\left[ n \right]} \right\}} } \right|^2}.
\end{align}
Submitting \eqref{g_ghat} and \eqref{g_ghat2} into \eqref{BU1}, we obtain
\begin{align}
  \mathbb{E}\left\{ {{{\left| {{\text{BU}}}\left[ n \right] \right|}^2}} \right\} = \kappa {p_{\text{u}}}\left[ n \right]\sum\limits_{l = 1}^L \left( {{\text{tr}}\left( {{{\mathbf{R}}_l}\left[ n \right]{{\mathbf{Q}}_l}\left[ n \right]} \right)} + {\mathbf{\bar h}}_l^{\text{H}}\left[ n \right]{{\mathbf{Q}}_l}\left[ n \right]{{{\mathbf{\bar h}}}_l}\left[ n \right] + {\mathbf{\bar h}}_l^{\text{H}}\left[ n \right]{{\mathbf{R}}_l}\left[ n \right]{{{\mathbf{\bar h}}}_l}\left[ n \right] \right).
\end{align}

\emph{3) Compute ${\mathbb{E}\left\{ {{{\left| {{\mathrm{HI}}}\left[ n \right] \right|}^2}} \right\}}$:}
Due to $\eta\left[ n \right]$ is independent with channels ${{\mathbf{g}}_l}\left[ n \right]$, we have
\begin{align}
  &\mathbb{E}\left\{ {{{\left| {{\text{HI}}}\left[ n \right] \right|}^2}} \right\} = \mathbb{E}\left\{ {{{\left| \eta\left[ n \right]  \right|}^2}} \right\}\mathbb{E}\left\{ {{{\left| {\sum\limits_{l = 1}^L {{\mathbf{\hat g}}_l^{\text{H}}\left[ n \right]{{\mathbf{g}}_l}\left[ n \right]} } \right|}^2}} \right\} \notag\\
 &= \left( {1 - \kappa } \right){p_{\text{u}}}\left[ n \right]\left( {\sum\limits_{l = 1}^L {\left( {{\text{tr}}\left( {{{\mathbf{R}}_l}\left[ n \right]{{\mathbf{Q}}_l}\left[ n \right]} \right) + {\mathbf{\bar h}}_l^{\text{H}}\left[ n \right]{{\mathbf{Q}}_l}\left[ n \right]{{{\mathbf{\bar h}}}_l}\left[ n \right] + {\mathbf{\bar h}}_l^{\text{H}}\left[ n \right]{{\mathbf{R}}_l}\left[ n \right]{{{\mathbf{\bar h}}}_l}\left[ n \right]} \right)} } \right. \notag \\
  &\left. { + {{\left( {\sum\limits_{l = 1}^L {\left( {{\text{tr}}\left( {{{\mathbf{Q}}_l}\left[ n \right]} \right) + {{\left\| {{{{\mathbf{\bar h}}}_l}\left[ n \right]} \right\|}^2}} \right)} } \right)}^2}} \right) .
\end{align}

\emph{4) Compute ${\mathbb{E}\left\{ {{{\left| {{\mathrm{NS}}}\left[ n \right] \right|}^2}} \right\}}$:}
Since the noise is independent with channels, we obtain
\begin{align}
  \mathbb{E}\left\{ {{{\left| {{\text{NS}}}\left[ n \right] \right|}^2}} \right\} = \mathbb{E}\left\{ {{{\left| {\sum\limits_{l = 1}^L {{\mathbf{\hat g}}_l^{\text{H}}\left[ n \right]{{\mathbf{n}}_l}}\left[ n \right] } \right|}^2}} \right\} = {\sigma ^2}\sum\limits_{l = 1}^L {\left( {{\text{tr}}\left( {{{\mathbf{Q}}_l}}\left[ n \right] \right) + {{\left\| {{{{\mathbf{\bar h}}}_l}}\left[ n \right] \right\|}^2}} \right)}.
\end{align}

\end{appendices}

\bibliographystyle{IEEEtran}
\bibliography{IEEEabrv,Ref}

\begin{thebibliography}{10}
\providecommand{\url}[1]{#1}
\csname url@samestyle\endcsname
\providecommand{\newblock}{\relax}
\providecommand{\bibinfo}[2]{#2}
\providecommand{\BIBentrySTDinterwordspacing}{\spaceskip=0pt\relax}
\providecommand{\BIBentryALTinterwordstretchfactor}{4}
\providecommand{\BIBentryALTinterwordspacing}{\spaceskip=\fontdimen2\font plus
\BIBentryALTinterwordstretchfactor\fontdimen3\font minus
  \fontdimen4\font\relax}
\providecommand{\BIBforeignlanguage}[2]{{%
\expandafter\ifx\csname l@#1\endcsname\relax
\typeout{** WARNING: IEEEtran.bst: No hyphenation pattern has been}%
\typeout{** loaded for the language `#1'. Using the pattern for}%
\typeout{** the default language instead.}%
\else
\language=\csname l@#1\endcsname
\fi
#2}}
\providecommand{\BIBdecl}{\relax}
\BIBdecl

\bibitem{zheng2021analysis}
J.~Zheng, J.~Zhang, and B.~Ai, ``{Analysis of UAV communication with power
  transfer under cell-free massive MIMO systems},'' \emph{Proc. IEEE ICC}, May
  2021.

\bibitem{wong2017key}
V.~W. Wong, R.~Schober, D.~W.~K. Ng, and L.-C. Wang, \emph{{Key Technologies
  for 5G Wireless Systems}}.\hskip 1em plus 0.5em minus 0.4em\relax Cambridge
  University Press, 2017.

\bibitem{ai20205g}
B.~Ai, A.~F. Molisch, M.~Rupp, and Z.-D. Zhong, ``{5G key technologies for
  smart railways},'' \emph{Proc. IEEE}, vol. 108, no.~6, pp. 856--893, Jun.
  2020.

\bibitem{zhang2020prospective}
J.~Zhang, E.~Bj{\"o}rnson, M.~Matthaiou, D.~W.~K. Ng, H.~Yang, and D.~J. Love,
  ``{Prospective multiple antenna technologies for beyond 5G},'' \emph{IEEE J.
  Sel. Areas Commun.}, vol.~38, no.~8, pp. 1637--1660, Aug. 2020.

\bibitem{Ngo2017Cell}
H.~Q. Ngo, A.~Ashikhmin, Y.~Hong, E.~G. Larsson, and T.~L. Marzetta,
  ``{Cell-free massive MIMO versus small cells},'' \emph{IEEE Trans. Wireless
  Commun.}, vol.~16, no.~3, pp. 1834--1850, Mar. 2017.

\bibitem{bjornson2019making}
E.~Bj{\"o}rnson and L.~Sanguinetti, ``{Making cell-free massive MIMO
  competitive with MMSE processing and centralized implementation},''
  \emph{IEEE Trans. Wireless Commun.}, vol.~19, no.~1, pp. 77--90, Jan. 2020.

\bibitem{ozdogan2019performance}
{\"O}.~{\"O}zdogan, E.~Bj{\"o}rnson, and J.~Zhang, ``{Performance of cell-free
  massive MIMO with Rician fading and phase shifts},'' \emph{IEEE Trans.
  Wireless Commun.}, vol.~18, no.~11, pp. 5299--5315, Nov. 2019.

\bibitem{chen2018channel}
Z.~Chen and E.~Bj{\"o}rnson, ``{Channel hardening and favorable propagation in
  cell-free massive MIMO with stochastic geometry},'' \emph{IEEE Trans.
  Commun.}, vol.~66, no.~11, pp. 5205--5219, Nov. 2018.

\bibitem{jin2019channel}
Y.~Jin, J.~Zhang, S.~Jin, and B.~Ai, ``{Channel estimation for cell-free mmWave
  massive MIMO through deep learning},'' \emph{IEEE Trans. Veh. Technol.},
  vol.~68, no.~10, pp. 10\,325--10\,329, Oct. 2019.

\bibitem{zheng2020efficient}
J.~Zheng, J.~Zhang, L.~Zhang, X.~Zhang, and B.~Ai, ``{Efficient receiver design
  for uplink cell-free massive MIMO with hardware impairments},'' \emph{IEEE
  Trans. Veh. Technol.}, vol.~69, no.~4, pp. 4537--4541, Apr. 2020.

\bibitem{9043712}
Y.~{Cai}, Z.~{Wei}, R.~{Li}, D.~W.~K. {Ng}, and J.~{Yuan}, ``{Joint trajectory
  and resource allocation design for energy-efficient secure UAV communication
  systems},'' \emph{IEEE Trans. Commun.}, vol.~68, no.~7, pp. 4536--4553, Jul.
  2020.

\bibitem{zeng2019accessing}
Y.~Zeng, Q.~Wu, and R.~Zhang, ``{Accessing from the sky: A tutorial on UAV
  communications for 5G and beyond},'' \emph{Proc. IEEE}, vol. 107, no.~12, pp.
  2327--2375, Dec. 2019.

\bibitem{zeng2016wireless}
Y.~Zeng, R.~Zhang, and T.~J. Lim, ``{Wireless communications with unmanned
  aerial vehicles: Opportunities and challenges},'' \emph{IEEE Commun. Mag.},
  vol.~54, no.~5, pp. 36--42, May 2016.

\bibitem{zeng2018cellular}
Y.~Zeng, J.~Lyu, and R.~Zhang, ``{Cellular-connected UAV: Potential,
  challenges, and promising technologies},'' \emph{IEEE Wireless Commun.},
  vol.~26, no.~1, pp. 120--127, Jan. 2018.

\bibitem{shen2020multi}
C.~Shen, T.-H. Chang, J.~Gong, Y.~Zeng, and R.~Zhang, ``{Multi-UAV interference
  coordination via joint trajectory and power control},'' \emph{IEEE Trans.
  Signal Process.}, vol.~68, pp. 843--858, Jan. 2020.

\bibitem{zhan2019completion}
C.~Zhan and Y.~Zeng, ``{Completion time minimization for multi-UAV-enabled data
  collection},'' \emph{IEEE Trans. Wireless Commun.}, vol.~18, no.~10, pp.
  4859--4872, Oct. 2019.

\bibitem{mozaffari2018beyond}
M.~Mozaffari, A.~T.~Z. Kasgari, W.~Saad, M.~Bennis, and M.~Debbah, ``{Beyond 5G
  with UAVs: Foundations of a 3D wireless cellular network},'' \emph{IEEE
  Trans. Wireless Commun.}, vol.~18, no.~1, pp. 357--372, Jan. 2018.

\bibitem{mei2019cellular}
W.~Mei, Q.~Wu, and R.~Zhang, ``{Cellular-connected UAV: Uplink association,
  power control and interference coordination},'' \emph{IEEE Trans. Wireless
  Commun.}, vol.~18, no.~11, pp. 5380--5393, Nov. 2019.

\bibitem{lyu2018uav}
J.~Lyu, Y.~Zeng, and R.~Zhang, ``{UAV-aided offloading for cellular hotspot},''
  \emph{IEEE Trans. Wireless Commun.}, vol.~17, no.~6, pp. 3988--4001, Jun.
  2018.

\bibitem{d2020analysis}
C.~D¡¯Andrea, A.~Garcia-Rodriguez, G.~Geraci, L.~G. Giordano, and S.~Buzzi,
  ``{Analysis of UAV communications in cell-free massive MIMO systems},''
  \emph{IEEE Open J. Commun. Society}, vol.~1, pp. 133--147, Jan. 2020.

\bibitem{d2019cell}
C.~D'Andrea, A.~Garcia-Rodriguez, G.~Geraci, L.~G. Giordano, and S.~Buzzi,
  ``{Cell-free massive MIMO for UAV communications},'' in \emph{Proc. IEEE
  ICC}, May 2019, pp. 1--6.

\bibitem{9336017}
J.~{An} and F.~{Zhao}, ``{Trajectory optimization and power allocation
  algorithm in MBS-assisted cell-free massive MIMO systems},'' \emph{IEEE
  Access}, pp. 1--1, 2021.

\bibitem{gupta2015survey}
L.~Gupta, R.~Jain, and G.~Vaszkun, ``{Survey of important issues in UAV
  communication networks},'' \emph{IEEE Commun. Surv. Tuts.}, vol.~18, no.~2,
  pp. 1123--1152, Nov. 2015.

\bibitem{mozaffari2019tutorial}
M.~Mozaffari, W.~Saad, M.~Bennis, Y.-H. Nam, and M.~Debbah, ``{A tutorial on
  UAVs for wireless networks: Applications, challenges, and open problems},''
  \emph{IEEE Commun. Surv. Tuts.}, vol.~21, no.~3, pp. 2334--2360, Mar. 2019.

\bibitem{liu2018energy}
C.~H. Liu, Z.~Chen, J.~Tang, J.~Xu, and C.~Piao, ``{Energy-efficient UAV
  control for effective and fair communication coverage: A deep reinforcement
  learning approach},'' \emph{IEEE J. Sel. Areas Commun.}, vol.~36, no.~9, pp.
  2059--2070, Sep. 2018.

\bibitem{wang2019coverage}
X.~Wang and M.~C. Gursoy, ``{Coverage analysis for energy-harvesting
  UAV-assisted mmWave cellular networks},'' \emph{IEEE J. Sel. Areas Commun.},
  vol.~37, no.~12, pp. 2832--2850, Dec. 2019.

\bibitem{xie2018throughput}
L.~Xie, J.~Xu, and R.~Zhang, ``{Throughput maximization for UAV-enabled
  wireless powered communication networks},'' \emph{IEEE Internet Things J.},
  vol.~6, no.~2, pp. 1690--1703, Feb. 2018.

\bibitem{zhou2018computation}
F.~Zhou, Y.~Wu, R.~Q. Hu, and Y.~Qian, ``{Computation rate maximization in
  UAV-enabled wireless-powered mobile-edge computing systems},'' \emph{IEEE J.
  Sel. Areas Commun.}, vol.~36, no.~9, pp. 1927--1941, Sep. 2018.

\bibitem{9075988}
W.~{Wang}, X.~{Li}, M.~{Zhang}, K.~{Cumanan}, D.~W. {Kwan Ng}, G.~{Zhang},
  J.~{Tang}, and O.~A. {Dobre}, ``{Energy-constrained UAV-assisted secure
  communications with position optimization and cooperative jamming},''
  \emph{IEEE Trans. Commun.}, vol.~68, no.~7, pp. 4476--4489, Jul. 2020.

\bibitem{hou2020hardware}
J.~Hou, Z.~Yang, and M.~Shikh-Bahaei, ``{Hardware impairment-aware data
  collection and wireless power transfer using a MIMO full-duplex UAV},'' in
  \emph{Proc. IEEE ICC}, May 2020, pp. 1--6.

\bibitem{li2020uav}
X.~Li, Q.~Wang, Y.~Liu, T.~A. Tsiftsis, Z.~Ding, and A.~Nallanathan,
  ``{UAV-aided multi-way NOMA networks with residual hardware impairments},''
  \emph{IEEE Wireless Commun. Lett.}, vol.~9, no.~9, pp. 1538--1542, Sep. 2020.

\bibitem{bjornson2017massive}
E.~Bj{\"o}rnson, J.~Hoydis, and L.~Sanguinetti, ``{Massive MIMO networks:
  Spectral, energy, and hardware efficiency},'' \emph{Foundations and
  Trends{\textregistered} in Signal Processing}, vol.~11, no. 3-4, pp.
  154--655, Nov. 2017.

\bibitem{mozaffari2017mobile}
M.~Mozaffari, W.~Saad, M.~Bennis, and M.~Debbah, ``{Mobile unmanned aerial
  vehicles (UAVs) for energy-efficient internet of things communications},''
  \emph{IEEE Trans. Wireless Commun.}, vol.~16, no.~11, pp. 7574--7589, Nov.
  2017.

\bibitem{zeng2018trajectory}
Y.~Zeng, X.~Xu, and R.~Zhang, ``{Trajectory design for completion time
  minimization in UAV-enabled multicasting},'' \emph{IEEE Trans. Wireless
  Commun.}, vol.~17, no.~4, pp. 2233--2246, Apr. 2018.

\bibitem{sun2019optimal}
Y.~Sun, D.~Xu, D.~W.~K. Ng, L.~Dai, and R.~Schober, ``{Optimal 3D-trajectory
  design and resource allocation for solar-powered UAV communication
  systems},'' \emph{IEEE Trans. Commun.}, vol.~67, no.~6, pp. 4281--4298, Jun.
  2019.

\bibitem{zhang2019securing}
G.~Zhang, Q.~Wu, M.~Cui, and R.~Zhang, ``{Securing UAV communications via joint
  trajectory and power control},'' \emph{IEEE Trans. Wireless Commun.},
  vol.~18, no.~2, pp. 1376--1389, Feb. 2019.

\bibitem{cui2018robust}
M.~Cui, G.~Zhang, Q.~Wu, and D.~W.~K. Ng, ``{Robust trajectory and transmit
  power design for secure UAV communications},'' \emph{IEEE Trans. Veh.
  Technol.}, vol.~67, no.~9, pp. 9042--9046, Sep. 2018.

\bibitem{gong2018flight}
J.~Gong, T.-H. Chang, C.~Shen, and X.~Chen, ``{Flight time minimization of UAV
  for data collection over wireless sensor networks},'' \emph{IEEE J. Sel.
  Areas Commun.}, vol.~36, no.~9, pp. 1942--1954, Sep. 2018.

\bibitem{yang2018energy}
D.~Yang, Q.~Wu, Y.~Zeng, and R.~Zhang, ``{Energy tradeoff in ground-to-UAV
  communication via trajectory design},'' \emph{IEEE Trans. Veh. Technol.},
  vol.~67, no.~7, pp. 6721--6726, Jul. 2018.

\bibitem{zhang2018uav}
J.~Zhang, Y.~Zeng, and R.~Zhang, ``{UAV-enabled radio access network:
  Multi-mode communication and trajectory design},'' \emph{IEEE Trans. Signal
  Process.}, vol.~66, no.~20, pp. 5269--5284, Oct. 2018.

\bibitem{xu2018uav}
J.~Xu, Y.~Zeng, and R.~Zhang, ``{UAV-enabled wireless power transfer:
  Trajectory design and energy optimization},'' \emph{IEEE Trans. Wireless
  Commun.}, vol.~17, no.~8, pp. 5092--5106, Aug. 2018.

\bibitem{srinidhi2011layered}
N.~Srinidhi, T.~Datta, A.~Chockalingam, and B.~S. Rajan, ``{Layered tabu search
  algorithm for large-MIMO detection and a lower bound on ML performance},''
  \emph{IEEE Trans. Commun.}, vol.~59, no.~11, pp. 2955--2963, Nov. 2011.

\bibitem{shaik2020mmse}
Z.~H. Shaik, E.~Bj{\"o}rnson, and E.~G. Larsson, ``{MMSE-optimal sequential
  processing for cell-free massive MIMO with radio stripes},''
  \emph{arXiv:2012.13928}, 2020.

\bibitem{ozdogan2019massive}
{\"O}.~{\"O}zdogan, E.~Bj{\"o}rnson, and E.~G. Larsson, ``{Massive MIMO with
  spatially correlated Rician fading channels},'' \emph{IEEE Trans. Commun.},
  vol.~67, no.~5, pp. 3234--3250, May 2019.

\end{thebibliography}

\end{document}